\documentclass{aastex}         
\usepackage{spr-astr-addons}
\usepackage{url}
\usepackage{graphicx,amssymb,amsmath,times}

\begin{document}
\title{Probing the evolution of the EBL photon density out to $z~\sim 1$ via $\gamma$-ray propagation measurements with \emph{Fermi}}

\shorttitle{Evolution of the EBL photon density via $\gamma$-ray propagation measurements}
\shortauthors{K. K. Singh et al.}

\author{K. K. Singh\altaffilmark{1}} 
\and 
\author{K. K. Yadav\altaffilmark{1,2}}
\and
\author{P. J. Meintjes\altaffilmark{3}}
\email{kksastro@barc.gov.in} 

\altaffiltext{1}{Astrophysical Sciences Division, Bhabha Atomic Research Centre, Mumbai- 400 085, India}
\altaffiltext{2}{Homi Bhabha National Institute, Mumbai- 400 094, India}
\altaffiltext{3}{Physics Department, University of the Free State, Bloemfontein - 9300, South Africa}

\begin{abstract}
The redshift ($z$) evolution of the Extragalactic Background Light (EBL) photon density is very important to 
understand the history of cosmological structure formation of galaxies and stars since the epoch of recombination. 
The EBL photons with the characteristic spectral energy distribution ranging from ultraviolet/optical to far-infrared 
provide a major source of opacity of the Universe to the GeV-TeV $\gamma$-rays travelling over cosmological distances. 
The effect of the EBL is very significant through $\gamma \gamma \rightarrow e^- e^+$ absorption process on the 
propagation of the $\gamma$-ray photons with energy $E >$ 50 GeV emitted from the sources at $z \sim 1$. This effect is 
characterized by the optical depth ($\tau$) which strongly depends on $E$, $z$ and density of the EBL photons. 
The proper density of the EBL photons increases with $z$ due to expansion of the Universe whereas evolution 
of radiation sources contributing to the EBL leads to a decrease in the density with increasing $z$. Therefore, 
the resultant volumetric evolution of the EBL photon density is approximated by a modified redshift dependence. 
In this work, we probe evolution of the EBL photon density predicted by two prominent models using cosmic gamma-ray 
horizon ($\tau (E,z)=$ 1) determined by the measurements from the \emph{Fermi}-Large Area Telescope (LAT) observations. 
The modified redshift dependence of the EBL photon density is optimized for a given EBL model by estimating the same gamma-ray 
horizon as predicted by the \emph{Fermi}-LAT observations. We further compare the optical depth estimates in the energy 
range $E =$ 4 GeV-1 TeV and redshift range $z =0.01-1$ from the \emph{Fermi}-LAT observations with the values 
derived from the two EBL models to further constrain the evolution of the EBL photon density in the $z~\sim 1$ Universe.
\end{abstract}

\keywords{cosmology: diffuse radiation, extragalactic background light: evolution, gamma-rays: general}

\section{Introduction}\label{s:intro}
The extragalactic background light (EBL) is the diffuse background radiation at ultraviolet (UV), optical, and infrared (IR) 
wavelengths. It is also described as the present epoch ($z = 0$) metagalactic radiation field associated with the star and 
galaxy formation \citep{Kneiske2002}. The dominant contributors to the EBL are direct starlight in the UV/optical wavelength 
band and reprocessed emission by dust in the host galaxies and interstellar medium in the IR waveband since the epoch of 
reionization \citep{Hauser2001,Dwek2013}. The spectral energy distribution (SED) of the EBL is observed to exhibit two 
distinct humps peaking at the optical and IR wavelengths and a valley between these two humps. Other radiation sources such 
as extremely faint galaxies, accretion on to the compact objects, active galactic nuclei, and  decay of the elementary particles 
can also contribute to the SED of the EBL \citep{Mattila2019}. The peak at IR wavelength may originate due to re-radiation from 
the significantly hotter dust in the torus of the active galactic nuclei. Therefore, intensity and spectrum of the EBL could 
provide important information about the nature of star formation, galaxy evolution, stellar and interstellar contents of the 
galaxies through the history of the Universe. The observed spectra of high redshift quasistellar sources suggest reionization 
of the intergalactic gas between the epoch of cosmic recombination ($z \approx 1100$) and 10$^9$ years later ($z \approx 6$) 
\citep{Fan2006}. The UV component of the EBL emitted by first stars galaxies is considered as the primary suspect for this 
process through photoionization \citep{Gilmore2009}. Thus, information about the UV radiation is very important to probe 
the phenomena of reionization in the early Universe \citep{Raue2012,Khaire2019,Cowley2019}. In general, understanding 
the properties of the broadband SED of the EBL photons is one of the attractive goals of the modern cosmology.
\par
Strict constraints on the intensity and SED of the EBL come mainly in three flavors: direct measurements, indirect measurements 
through the high energy $\gamma$-ray observations, and estimations from the integrated galaxy counts from the resolved 
source populations. Direct measurements of the EBL intensity are subject to very large uncertainties due to strong foreground 
emissions from the solar system, interplanetary dust (zodiacal light) and Milky-Way (diffuse galactic light) in the same 
wavelength band \citep{Hauser2001,Hauser1998}. Recent attempts for direct measurement of the EBL intensity as a function of 
wavelength are found to be very challenging and limited by the systematic uncertainties \citep{Matsuura2017,Zemcov2017}. 
An alternative method for indirect measurement of the EBL involves the effect of $\gamma-\gamma$ absorption via pair production 
during propagation of the high energy $\gamma$-ray photons emitted from the sources at the cosmological distances \citep{Gould1966,Stecker1992}. 
This method is also challenged by the set of uncertainties related to the measurement and determination of the spectra of distant 
$\gamma$-ray emitters. However, several stringent upper limits have been derived on the intensity of the EBL by $\gamma$-ray observations 
of distant blazars and assuming different spectral forms for their intrinsic spectra \citep{Aharonian2006,Mazin2007,Meyer2012,Singh2014,
Singh2019,Singh2020}. Integral of the light emitted by all resolved galaxies provides a strict lower limit to the EBL 
intensity \citep{Madau2000,Dole2006,Keenan2010,Driver2016}. Several promising models for the SED of the EBL at $z = 0$ have been proposed 
using different distinct approaches based on the above constraints \citep{Kneiske2010,Finke2010,Dom2011,Gilmore2012,Stecker2016,Franceschini2017}. 
Most of these EBL models are found to be in good agreement with the lower limits from  the resolved galaxy counts and scaling of a few 
of them combined with the high energy $\gamma$-ray observations provide well defined measurements of the EBL intensity at the present 
epoch \citep{Ackermann2012,Abramowski2013,Biteau2015,Ahnen2016a,Desai2019,Abeysekara2019}. Recently, analysis of the high energy $\gamma$-ray photons 
emitted from the active galaxies and detected by the \emph{Fermi}-Large Area Telescope (LAT) has been used to determine the intensity of the 
EBL up to redshift $z \sim 6$, i.e. light emission over 90$\%$ of the cosmic time \citep{Abdollahi2018}. The EBL spectrum determined 
by the \emph{Fermi}-LAT at the present epoch ($z = 0$) is consistent with the predictions from the method of the 
resolving individual galaxies. 
\par
The EBL intensity has also been constrained from the measurements of the $\gamma$-ray attenuation effects on the GeV-TeV spectra 
of the blazars observed with the current generation ground-based atmospheric Cherenkov telescopes like VERITAS, H.E.S.S. and MAGIC 
up to redshifts $z \sim$ 1. Observation of the $\gamma$-ray emission up to $\sim$ 200 GeV from the blazar PKS 1441+25 at 
$z = 0.939$ with the VERITAS telescopes has set a stringent upper limit on the EBL intensity broadly consistent with the 
resolved galaxies surveys \citep{Abeysekara2015}. This has provided an excellent baseline with the redshifted UV emission from 
the primordial stars. Recent model-independent measurement of the EBL from the $\gamma$-ray spectra of 14 VERITAS-detected 
blazars at $z = 0.044-0.604$ also shows good agreement with the lower limits derived from the resolved galaxies 
counts \citep{Abeysekara2019}. The H.E.S.S. collaboration also derived a model-independent SED of the EBL using the $\gamma$-ray observations 
of a sample of blazars in the redshift range $z = 0.031-0.287$ \citep{Abdalla2017}. The EBL intensity levels extracted in the different 
spectral bands are found to be in line with the results obtained from the \emph{Fermi}-LAT measurements close to the lower limits in the 
optical range \citep{Ackermann2012} and are also consistent with the upper limits derived from the VERITAS observations \citep{Abeysekara2015}. 
The MAGIC collaboration presented EBL constraints based on a joint likelihood analysis of 32 $\gamma$-ray spectra for 12 blazars in the 
redshift range $z = 0.031-0.944$ obtained by the MAGIC telescopes and the \emph{Fermi}-LAT \citep{Acciari2019}. A wavelength-resolved 
determination of the EBL indicated an excess in the UV/optical component of the SED relative to other models. However, this is 
compatible with the existing EBL models within statistical uncertainties. At high redshifts, the $\gamma$-ray bursts offer a significant 
advantage over the blazars for constraining the EBL intensity. Analysis of a sample of 22  $\gamma$-ray bursts detected by 
the \emph{Fermi}-LAT in the energy range of 65 MeV-500 GeV has been used to place first constraint on the UV component of the EBL 
at an effective redshift $z \sim 1.8$ \citep{Desai2017}.
\par
In the present work, we study the redshift evolution of the proper density of the EBL photons in the local Universe $z \le 1$ 
using predictions from the $\gamma$-ray observations with the \emph{Fermi}-LAT. We have used two most promising and widely used 
SEDs of the EBL at $z = 0$ proposed by \cite{Finke2010} and \cite{Dom2011} to probe the EBL evolution at lower redshifts. 
We first discuss the cosmological evolution of the EBL in Section \ref{s:ebl-evol}. Propagation of the $\gamma$-rays in the Universe 
and recent predictions from the \emph{Fermi}-LAT observations are described in Section \ref{s:gamma-prop}. In Section \ref{s:res-disc}, 
we present and discuss the results followed by the conclusion of this study in Section \ref{s:conclu}.

\begin{figure}[tb]
\includegraphics[width=0.7\columnwidth,angle=-90]{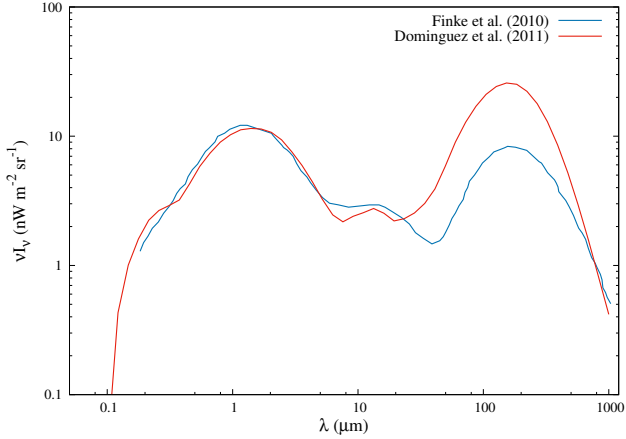}
\caption{Spectral energy distribution of the EBL at the present epoch ($z = 0$) proposed by Finke et al. (2010) and  Dom\'inguez et al. (2011)}
\label{ebl-sed}
\end{figure}

\section{EBL Evolution}\label{s:ebl-evol}
Accelerated expansion of the Universe has been confirmed and very well understood by the observations of type Ia 
supernovae \citep{Riess1998,Perlmutter1999}. The dynamics of expanding Universe is described by a free function 
of time called the scale factor $a(t)$, which is expressed as 
\begin{equation}\label{scale-fact}
	a(t)~=~a_0 (1 + z)^{-1}
\end{equation}
where $a_0$ is the scale factor at the present epoch corresponding to $z = 0$. The comoving radial distance is 
proportional to $a(t)$ and therefore the density of the EBL photons (number of photons per unit volume) evolves as 
\begin{equation}\label{volumetric}
	n(z)~\propto~(1 + z)^3 
\end{equation}
This implies that the photon proper density increases with redshift due to the expansion of the Universe. This is generally 
referred to as the volumetric evolution of the background photons. An observer in a galaxy at redshift $z > 0$ would observe 
a Universe which is smaller than the present day Universe by a factor $(1 + z)^3$. Since, the EBL represents integrated 
cosmic activities involving star and galaxy formation and models of dust or matter distribution in the galaxies, it is 
very important to consider the evolution of radiation sources contributing to the EBL intensity. Thus, at any given epoch, 
the proper number density of the EBL photons consists of accumulated radiation emitted at the previous epochs and 
their sources in the rest frame. During most of the cosmic time to the present epoch, the stars and galaxies progressively 
emit photons contributing to the EBL. Sources contributing to the optical regime of the EBL are at lower redshift 
($ z \le 0.6$) whereas IR photons originate at higher redshifts ($ z > 0.6$). This implies that increase in the proper 
photon density with redshift is larger for the IR and smaller for the optical photons. The enhancement in the proper photon 
density of the optical photons due to the volumetric evolution (Equation \ref{volumetric}) can be quickly compensated by 
the decrease in the population of available photons with the increasing redshift. Therefore, the effective comoving density of 
photons decreases at larger redshifts. To account for this, an evolutionary parameter $k$ is introduced to scale the proper number 
density of the EBL photons as 
\begin{equation}
	n(z)~\propto~(1 + z)^{3 - k}
\end{equation}
The value of $k$ can vary with redshift as it quantifies the effect of radiation sources contribution to the EBL. It plays a 
very important role in the propagation of the high energy $\gamma$-ray photons over cosmological distances. There is no 
uniquely determined value of $k$ and multiple values are proposed in the literature \citep{Madau1996,Aharonian2007,Raue2008}. 
In case of no radiation source, $k = 0$ indicates strong evolution of the optical emission of the galaxies with no absorption 
or reprocessing and photons are already present at the given redshift. A significant amount of the UV photons emitted at the 
early epochs are redshifted to the optical due to expansion of the Universe. In the case of the static Universe, the photon 
number density is higher than that integrated over redshift. Therefore, evolution of the galaxies should be properly considered 
while estimating the cosmological dependence of the number density of the EBL photons. The comoving number density of the EBL photons 
in the energy range $\epsilon$ and $\epsilon$ + d$\epsilon$ at redshift $z$ is given by 
\begin{equation}\label{ebl-den}
	n(\epsilon,z)~=~n(\epsilon_0,0) (1 + z)^{3 - k}
\end{equation}
where $\epsilon_0 = \epsilon (1 +z)^{-1}$ is the observed energy of the EBL photon at $z = 0$. The  comoving number density 
at the present epoch can be estimated from the intensity ($\nu I_\nu$) of the EBL using the relation \citep{Dwek2013}
\begin{equation}
     n(\epsilon_0,0)~[\rm ph~cm^{-3}~eV^{-1}]~=~\frac{4\pi}{c} \frac{\nu I_\nu (\nu_0,0)[\rm nW~m^{-2}~sr^{-1}]}{\epsilon_0^2 [\rm eV]} 
\end{equation}
where $\nu$ and $\nu_0$ are the frequencies corresponding to $\epsilon$ and  $\epsilon_0$ respectively. The broadband SED of the EBL 
is represented by $\nu I_\nu$ vs $\lambda$ (wavelength) on log-log scale as shown in Figure \ref{ebl-sed} for the two widely 
used models described in \citep{Finke2010,Dom2011}. The model proposed by \cite{Finke2010} assumes main-sequence stars as 
blackbodies which re-emit the star light absorbed by the dust after taking into account the star formation rate, initial mass function 
and dust extinction. It also includes emission from the post-main-sequence stars to model the broadband SED of the EBL photons which 
is very close to the lower limits from the galaxy counts at $z = 0$. The second model by  \cite{Dom2011} is based on the multi-wavelength 
data of about 6000 galaxies from different surveys and the rest frame K-band galaxy luminosity function which provides an accurate 
measurement of the galaxy evolution. Recently, a new determination of the evolving SED of the EBL up to $z \sim 6$ purely based on 
the deepest multi-wavelength observations from the UV to the far-IR of more than 150,000 galaxies has been reported \citep{Saldana2020}. 
The UV/optical peak of the SED derived in this new model for $z \le 1$ is compatible with the Finke et al. (2010) and  
Dom\'inguez et al. (2011) models. However, there is a large disagreement beween these models at all redshifts in the IR range. 
In this work, we adopt the broadband SED of the EBL at $z = 0$ predicted by above two models (shown in Figure \ref{ebl-sed}) to probe the 
evolution of the EBL photon density in the local Universe ($z \le 1$) using the propagation of high energy $\gamma$ rays emitted 
at different redshifts.

\begin{figure}[tb]
\includegraphics[width=0.7\columnwidth,angle=-90]{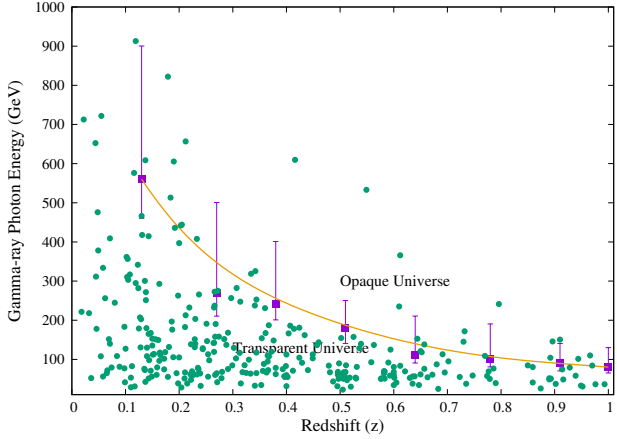}
\caption{Orange Curve: Gamma Ray Horizon from the \emph{Fermi}-LAT observations (purple filled squares) as a function of 
	redshift up to $z~\sim$ 1. Green Filled Circles: Highest energy of the $\gamma$-ray photons detected from the blazars 
	used in the estimation of the EBL density by the \emph{Fermi}-LAT Collaboration \citep{Abdollahi2018} 
	(Data and materials available online)}
\label{lat-horizon}
\end{figure}

\begin{figure}[tb]
\includegraphics[width=0.7\columnwidth,angle=-90]{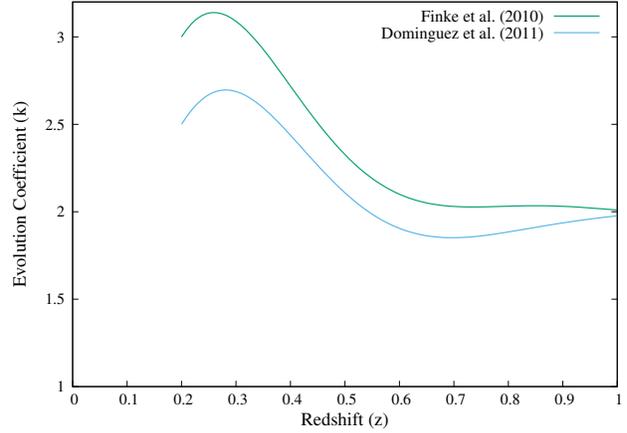}
\caption{Optimization of the evolution coefficient $k$ using the \emph{Gamma Ray Horizon} from the 
	\emph{Fermi}-LAT observations in the energy range $E_0 \approx$ 100-500 GeV and redhsift range $z =$ 0.2-1}
\label{k-hor-opt}
\end{figure}

\begin{figure}[tb]
\includegraphics*[width=0.32\columnwidth,angle=-90]{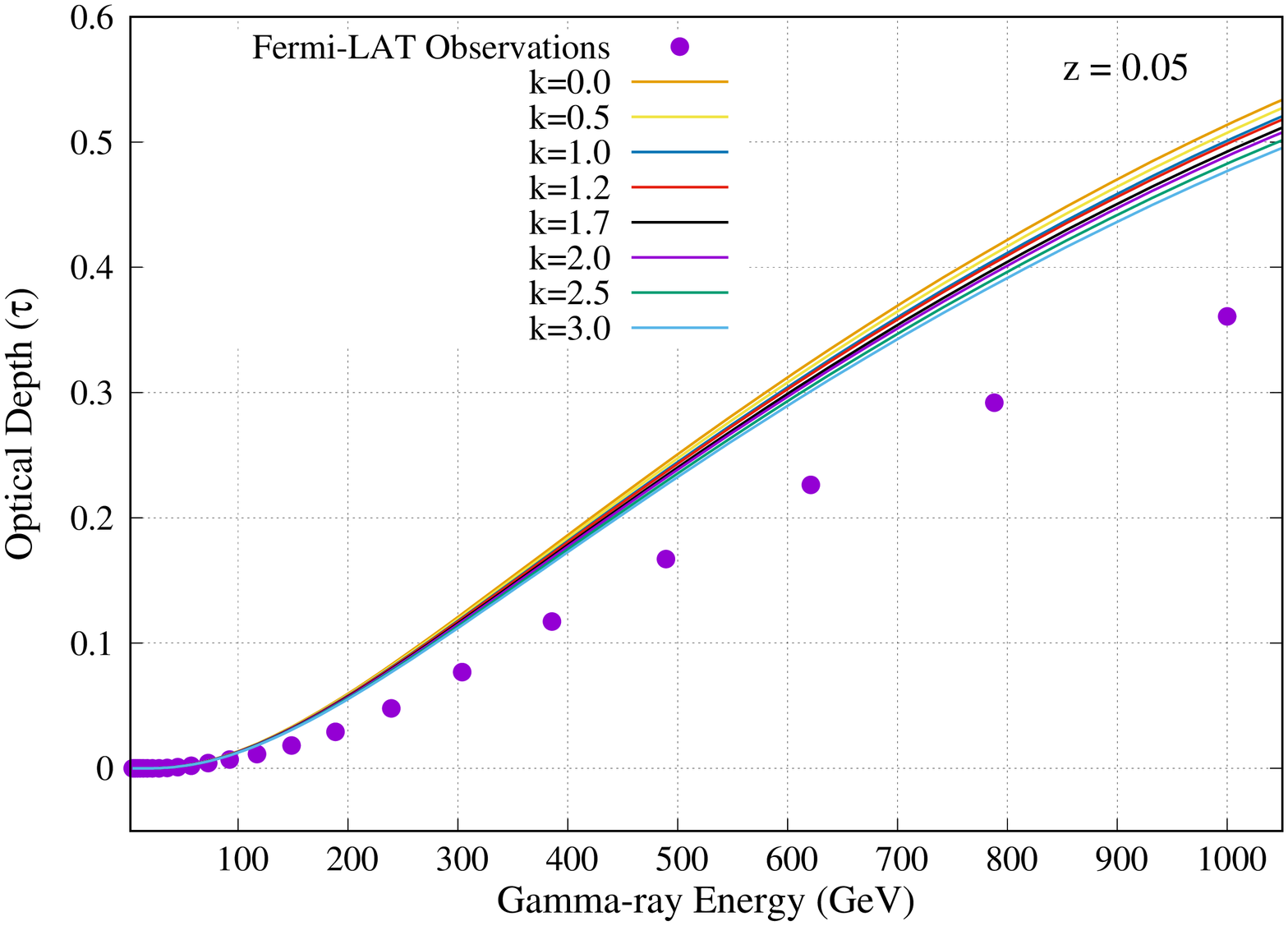}
\includegraphics*[width=0.32\columnwidth,angle=-90]{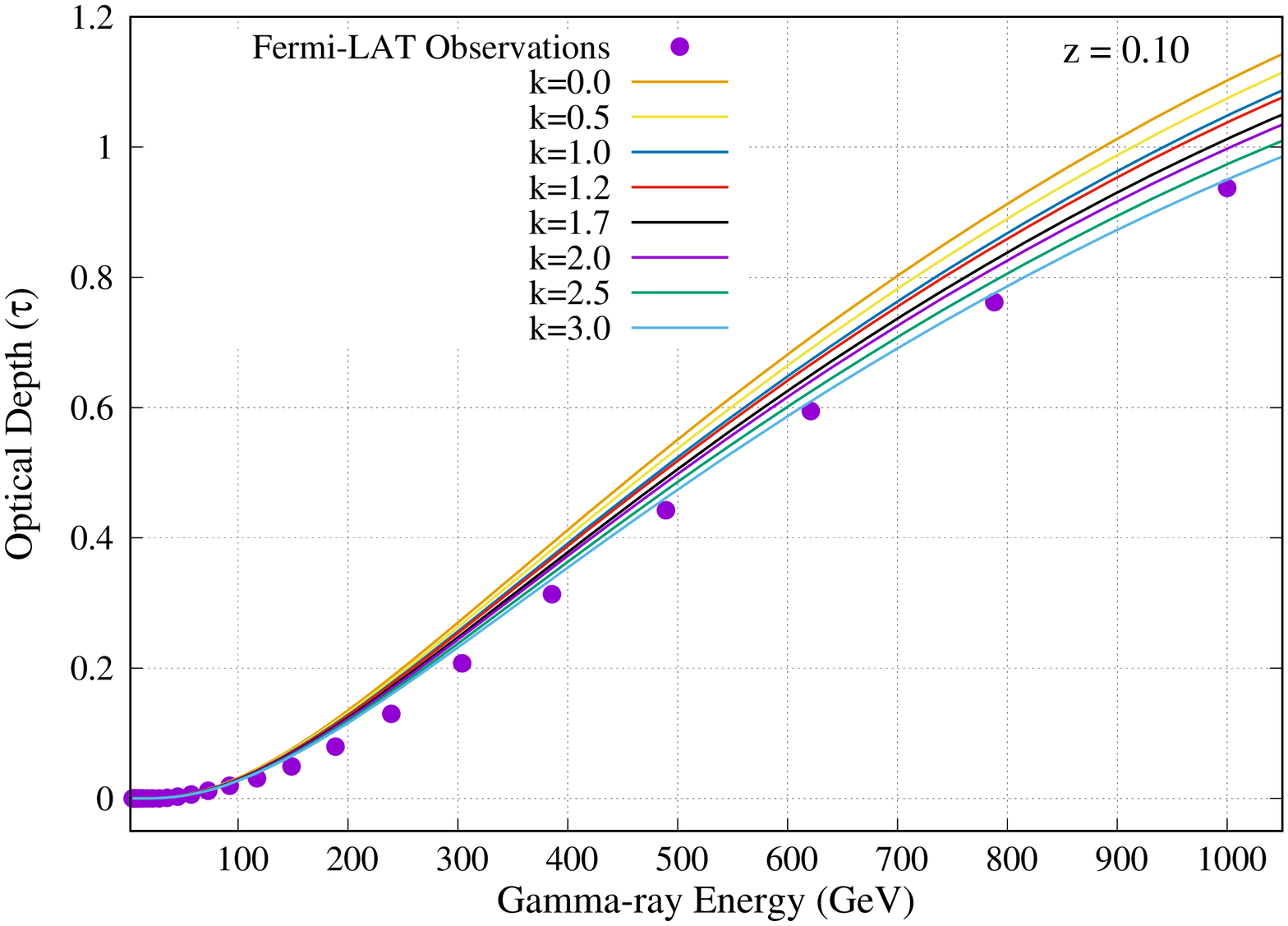}
\includegraphics*[width=0.32\columnwidth,angle=-90]{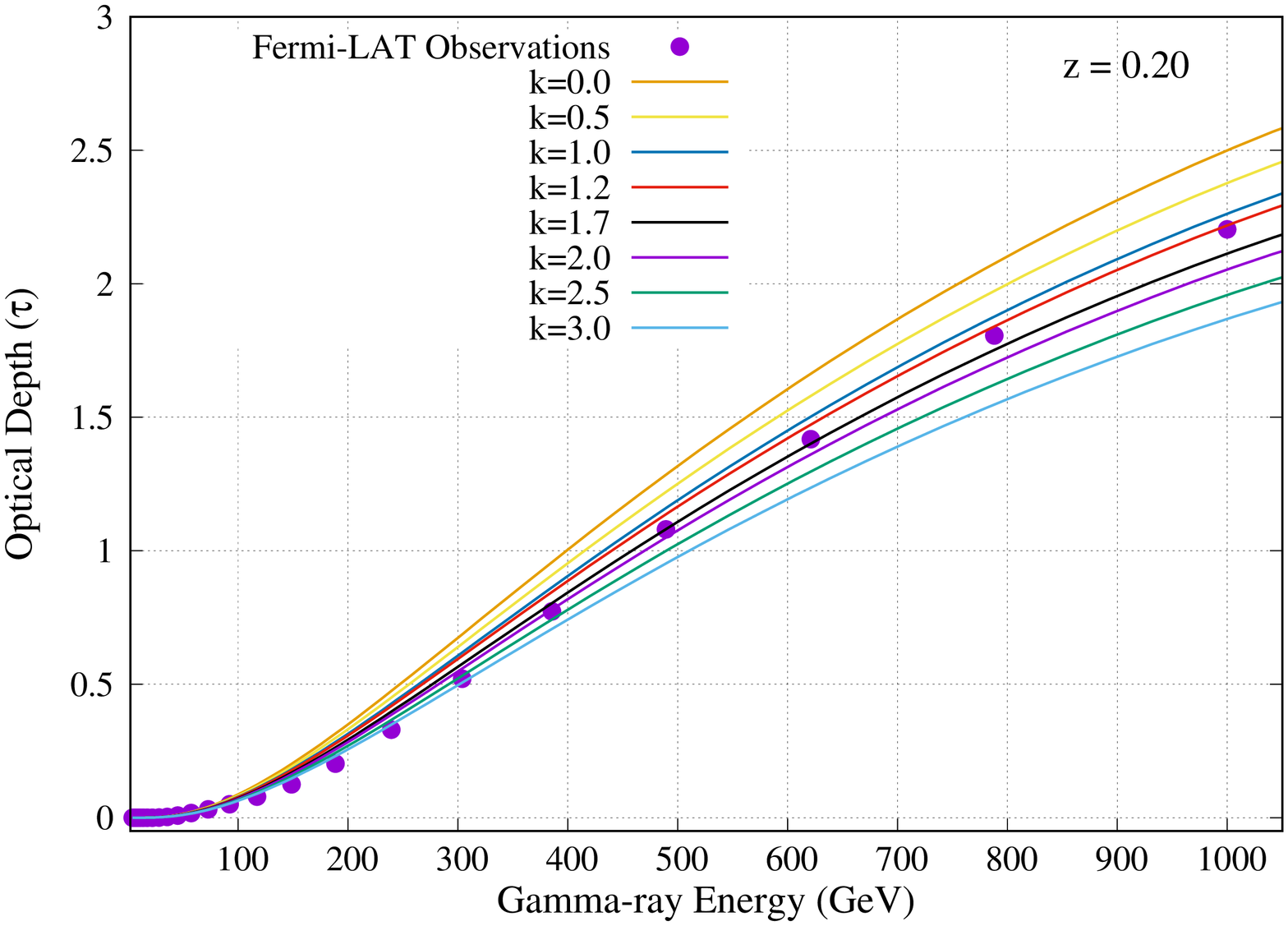}
\includegraphics*[width=0.32\columnwidth,angle=-90]{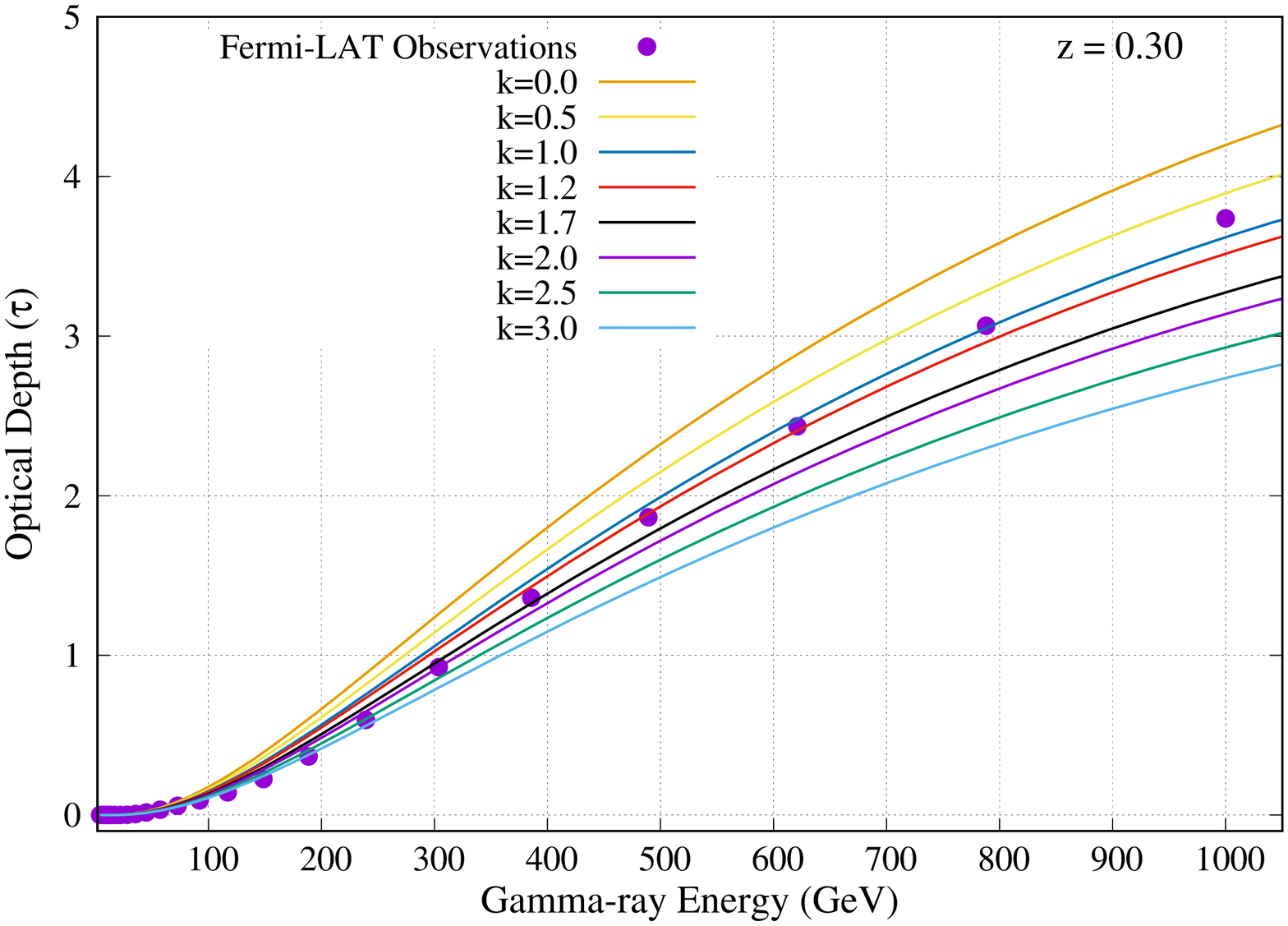}
\includegraphics*[width=0.32\columnwidth,angle=-90]{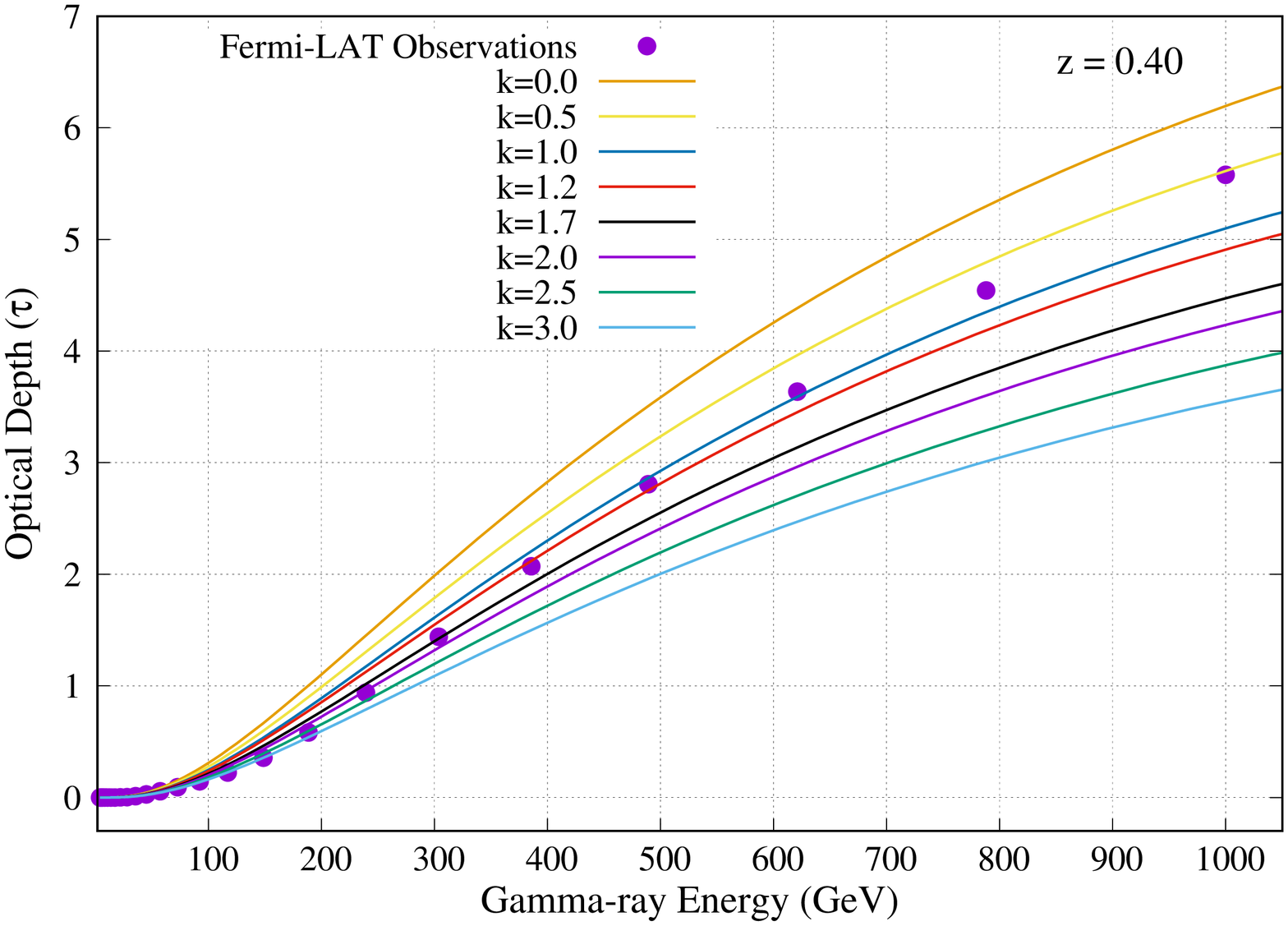}
\includegraphics*[width=0.32\columnwidth,angle=-90]{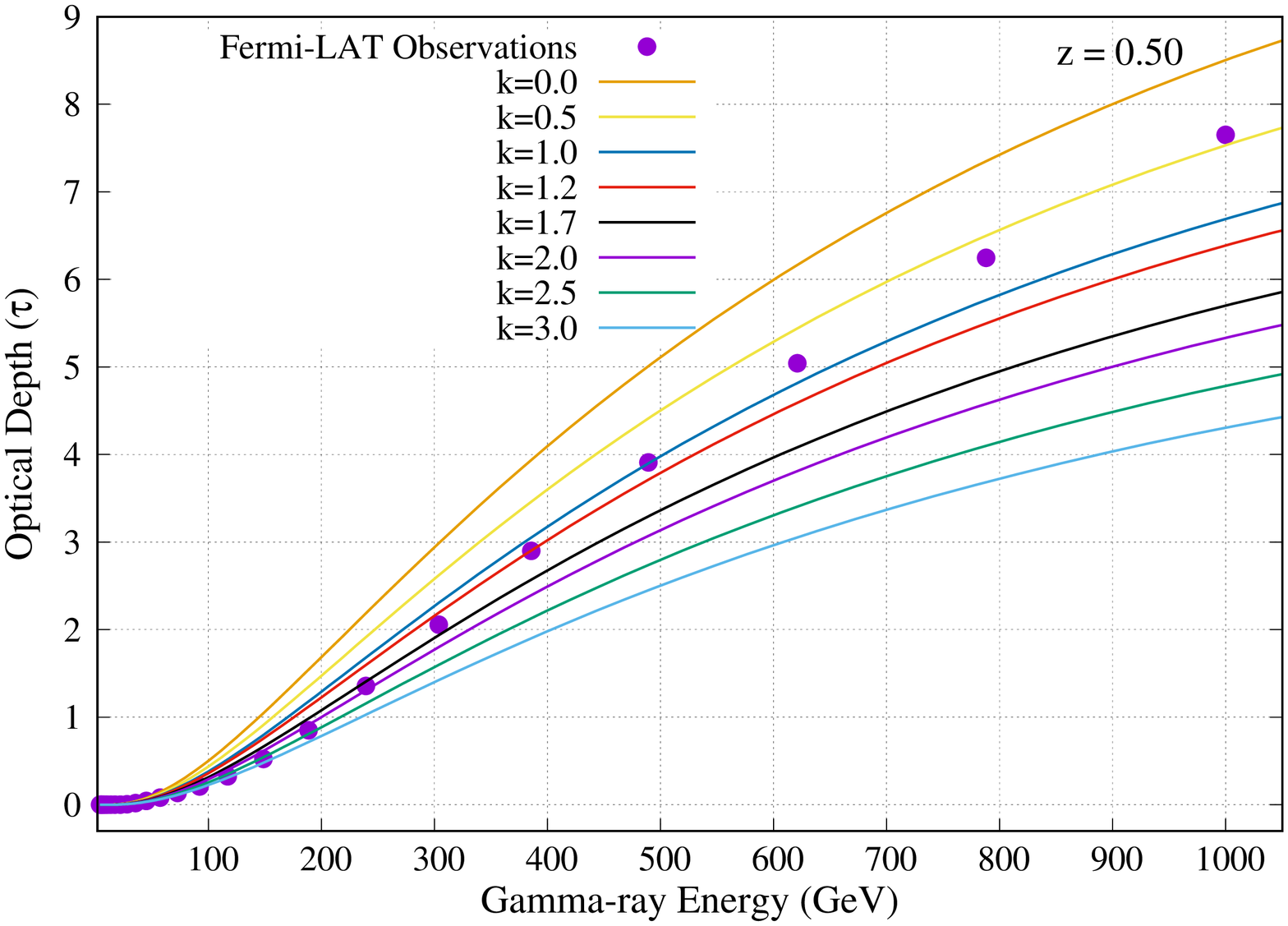}
\includegraphics*[width=0.32\columnwidth,angle=-90]{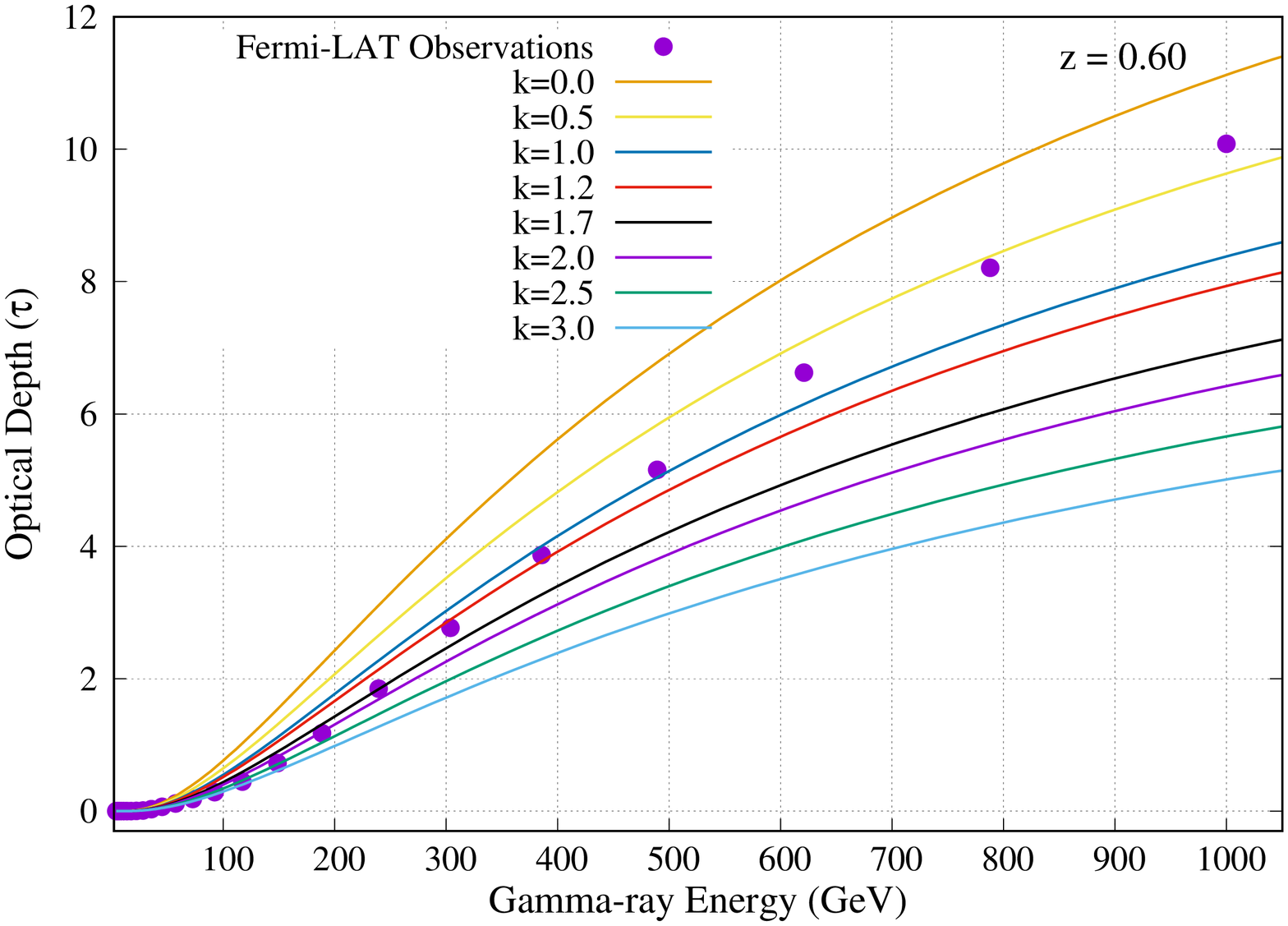}
\includegraphics*[width=0.32\columnwidth,angle=-90]{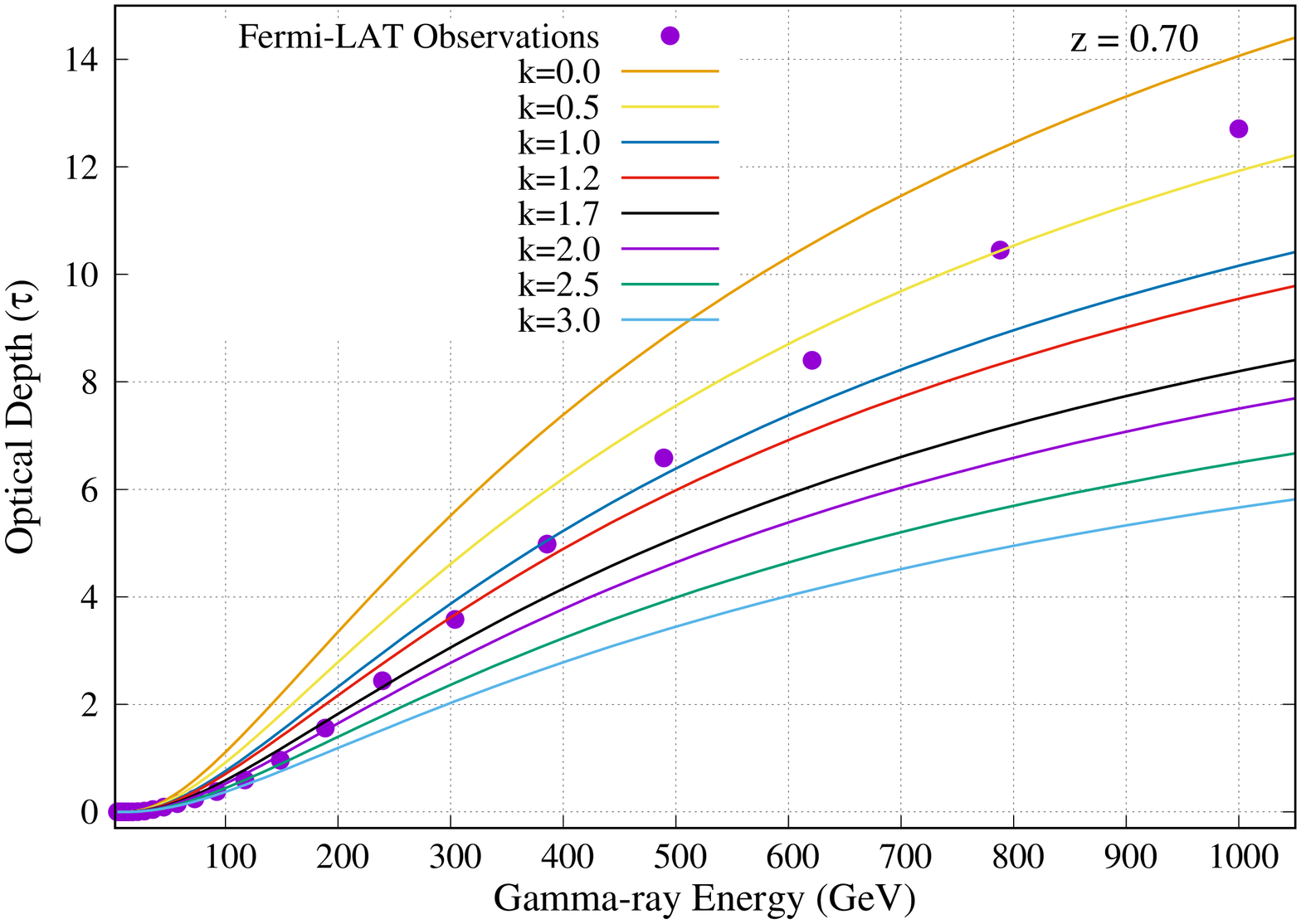}
\includegraphics*[width=0.32\columnwidth,angle=-90]{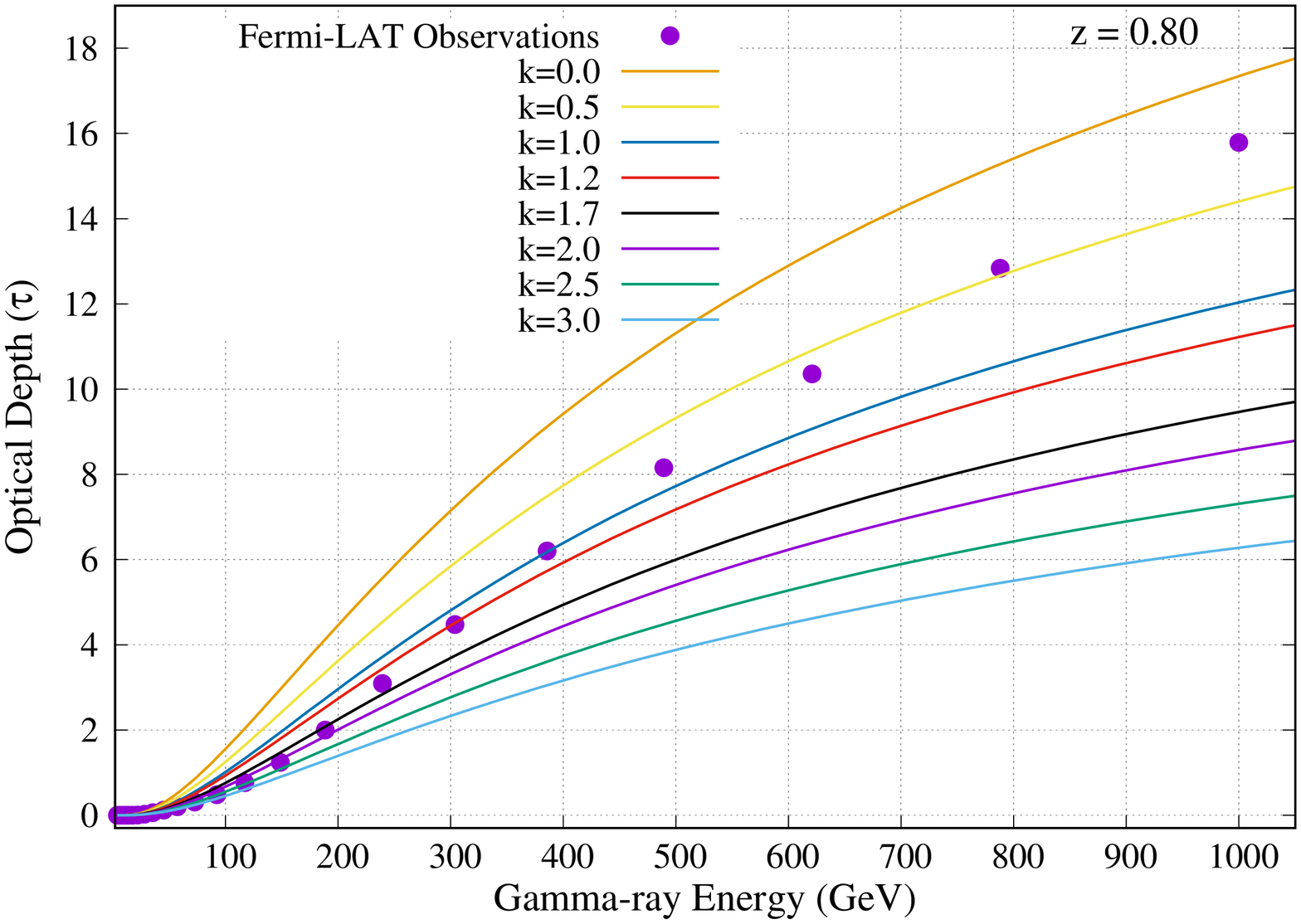}
\includegraphics*[width=0.32\columnwidth,angle=-90]{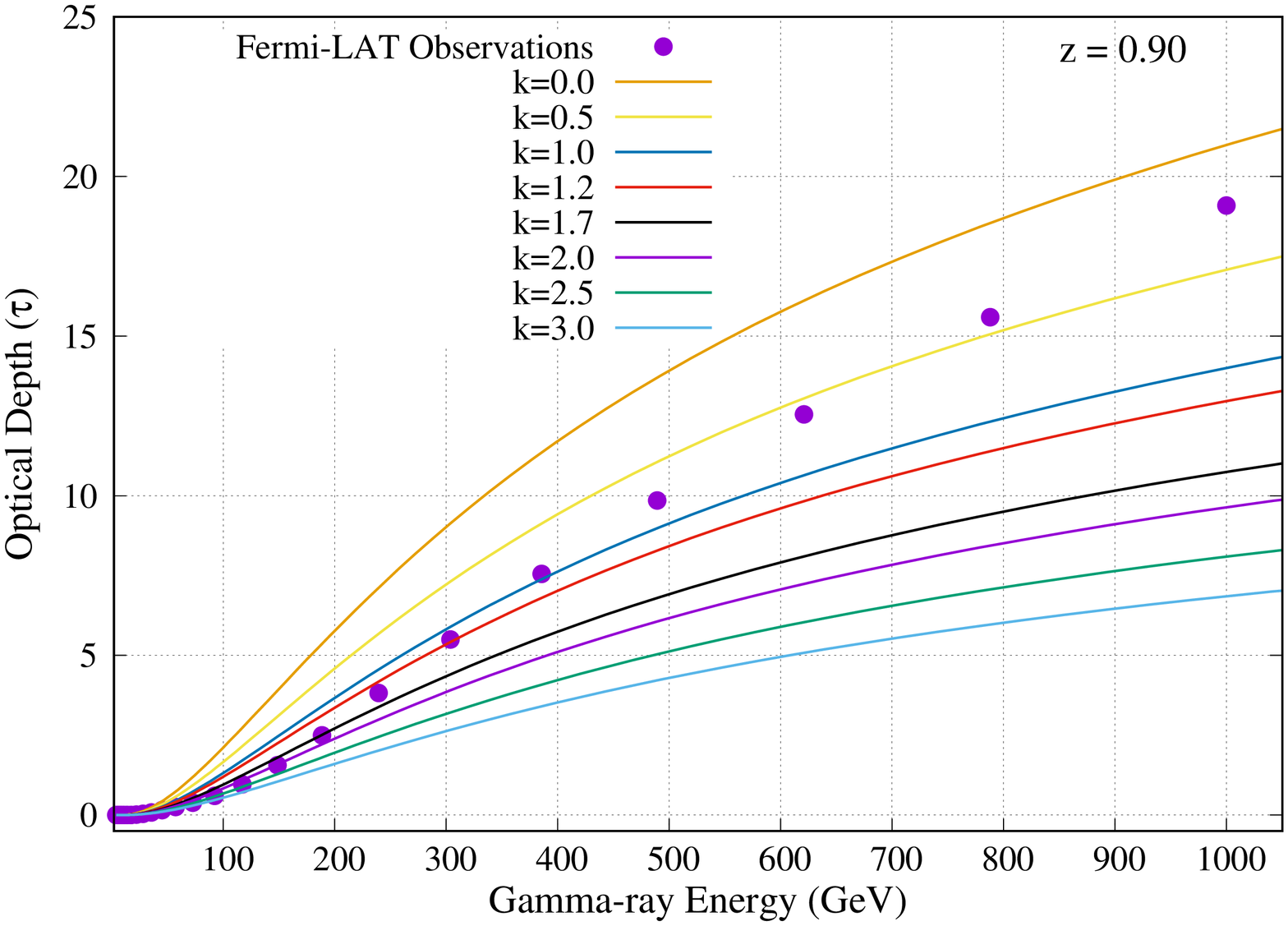}
\includegraphics*[width=0.32\columnwidth,angle=-90]{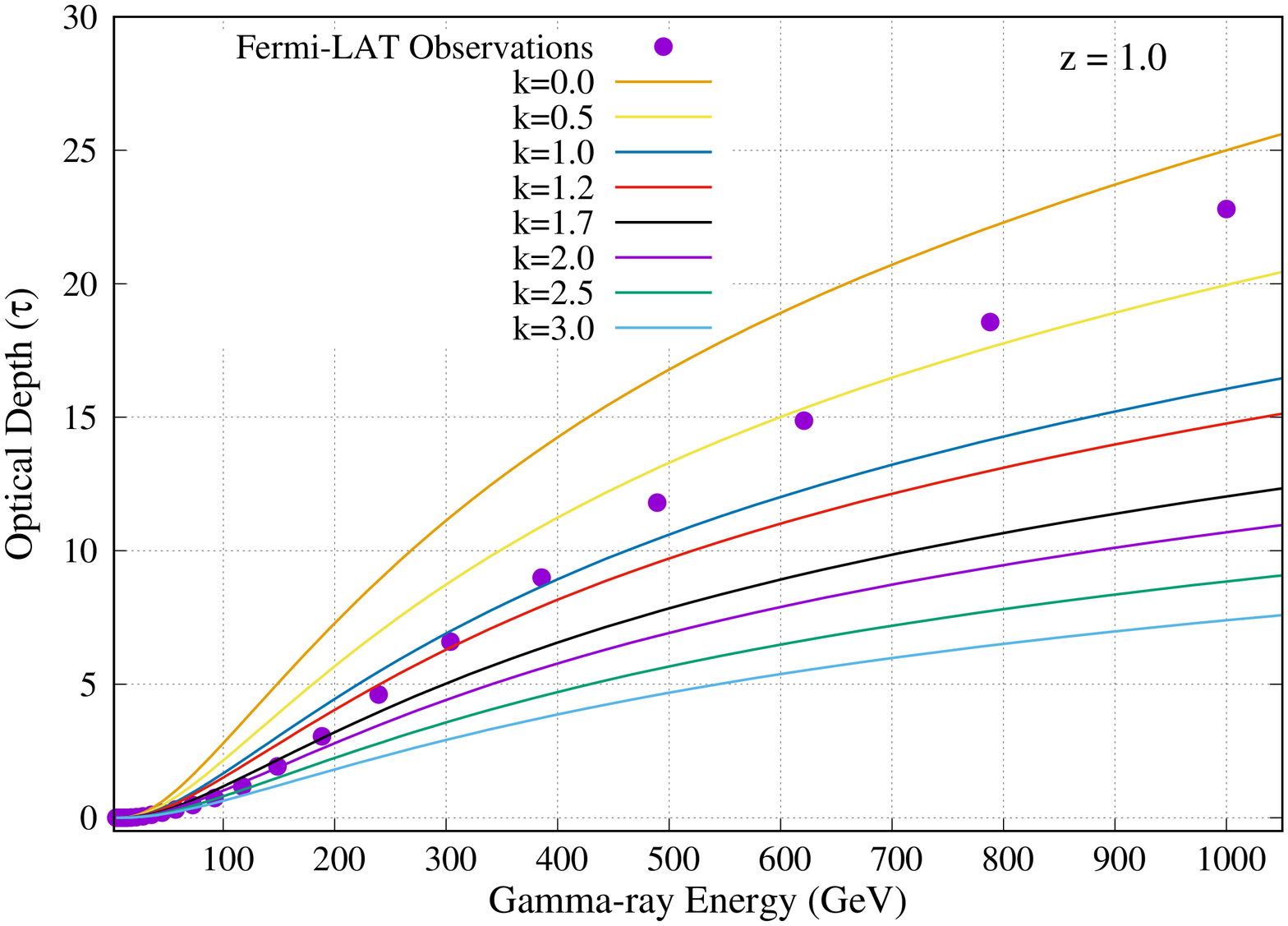}
\caption{Comparison of the optical depth values calculated using  the Finke et al. (2010) model for different values of $k$ at various 
	  redshifts with the corresponding opacity estimates derived using the EBL model proposed by \cite{Abdollahi2018} from the 
	  \emph{Fermi}-LAT observations}
\label{od-Finke}
\end{figure}

\begin{figure}[tb]
\includegraphics*[width=0.32\columnwidth,angle=-90]{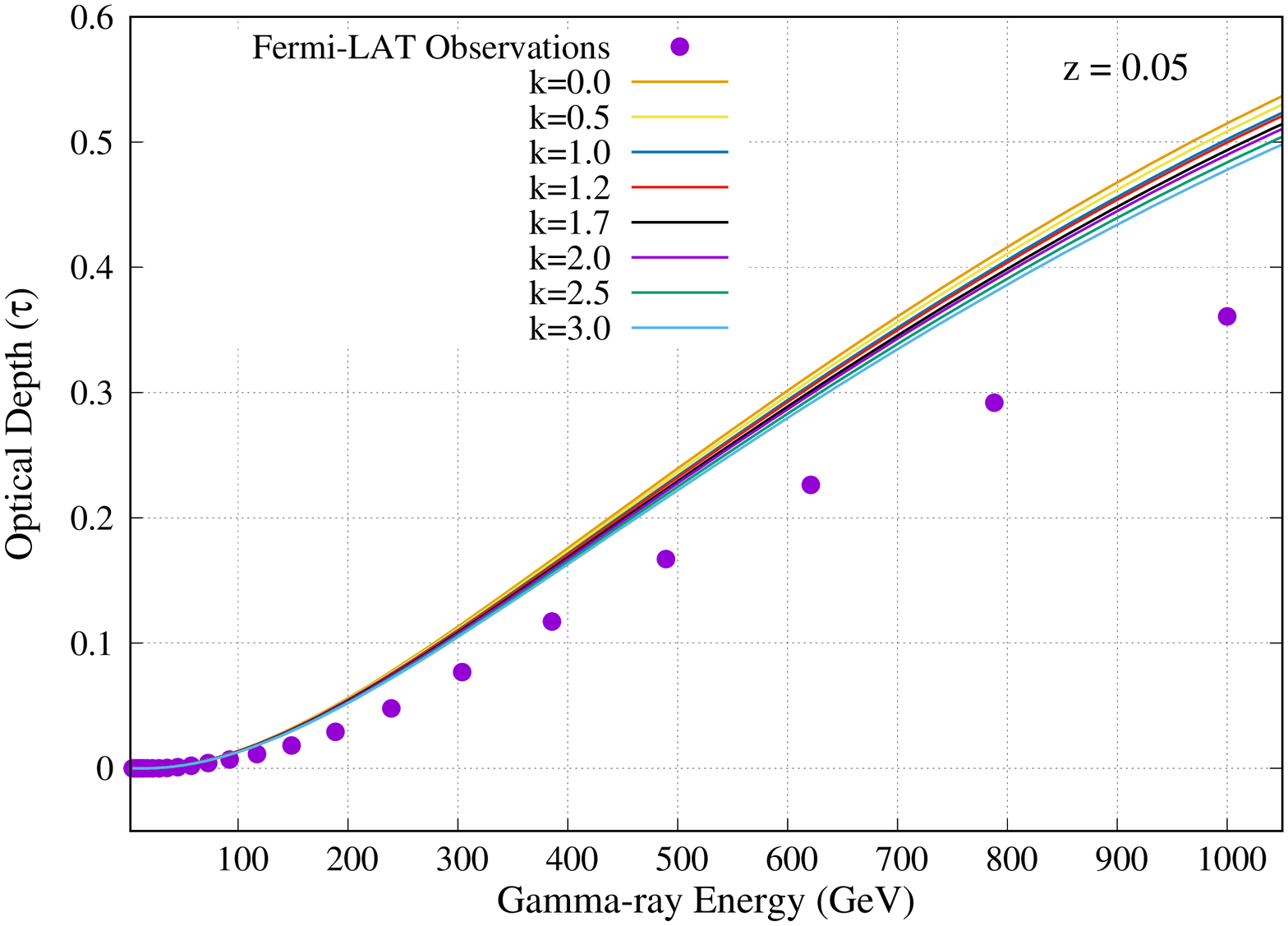}
\includegraphics*[width=0.32\columnwidth,angle=-90]{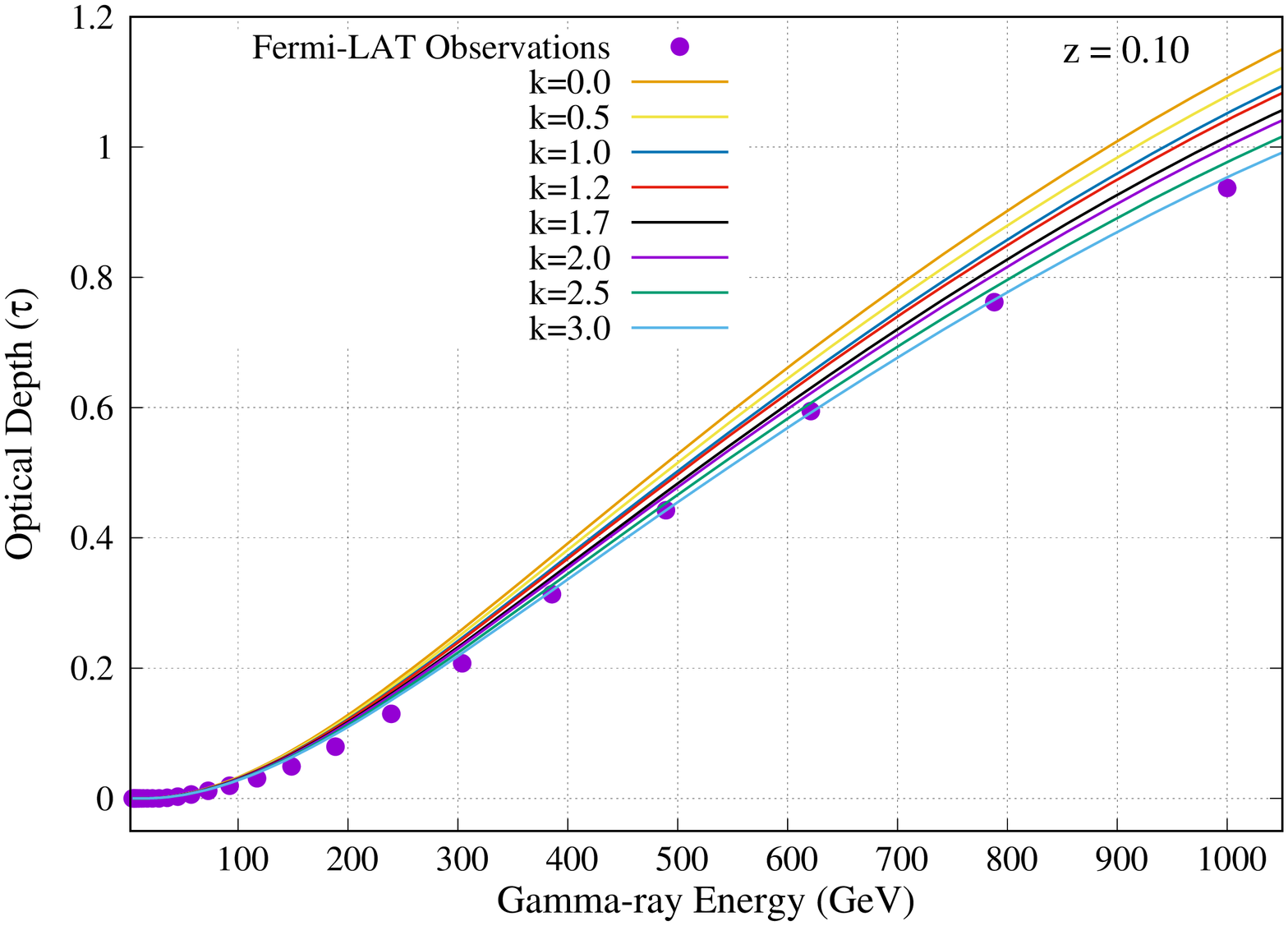}
\includegraphics*[width=0.32\columnwidth,angle=-90]{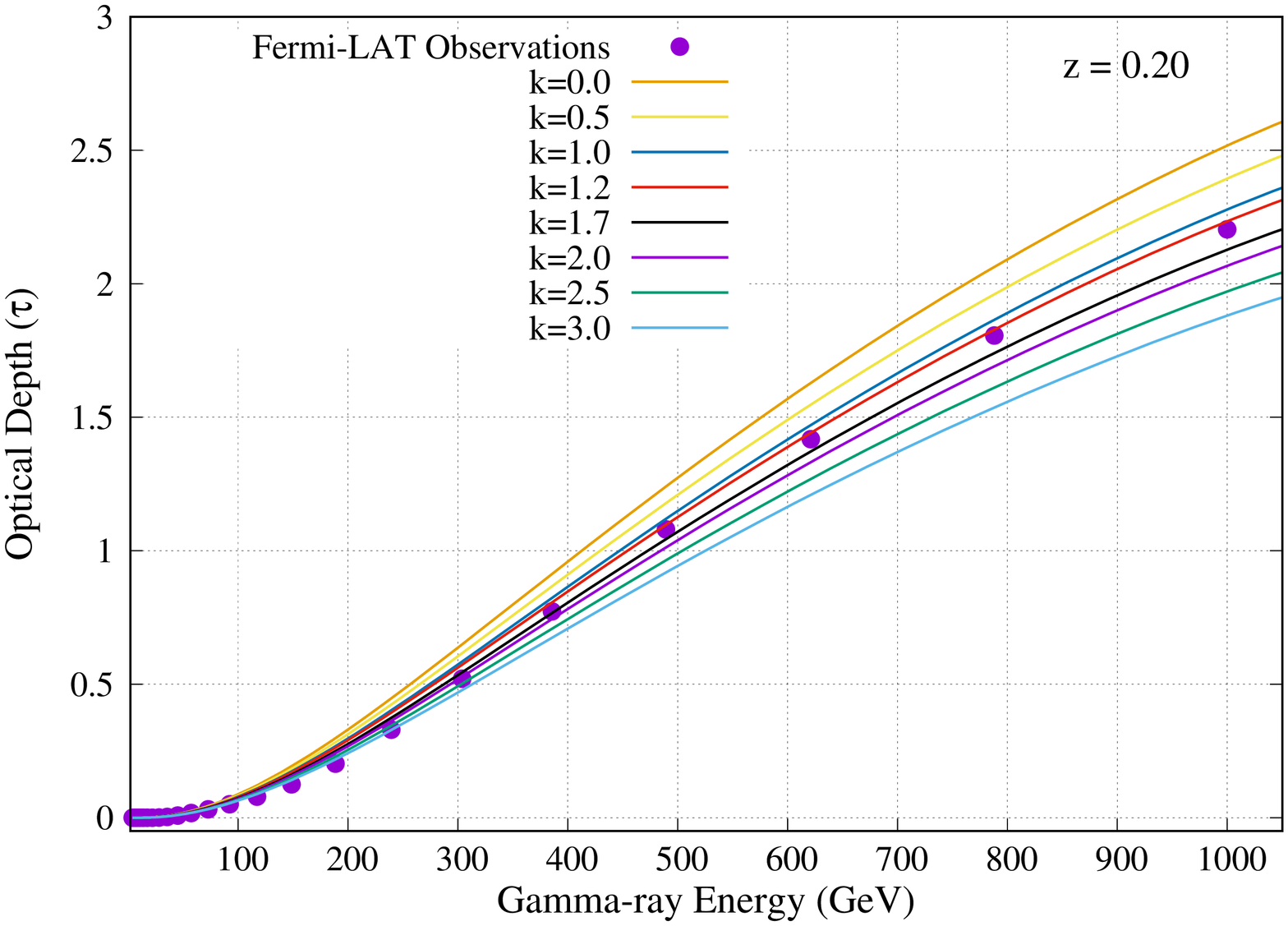}
\includegraphics*[width=0.32\columnwidth,angle=-90]{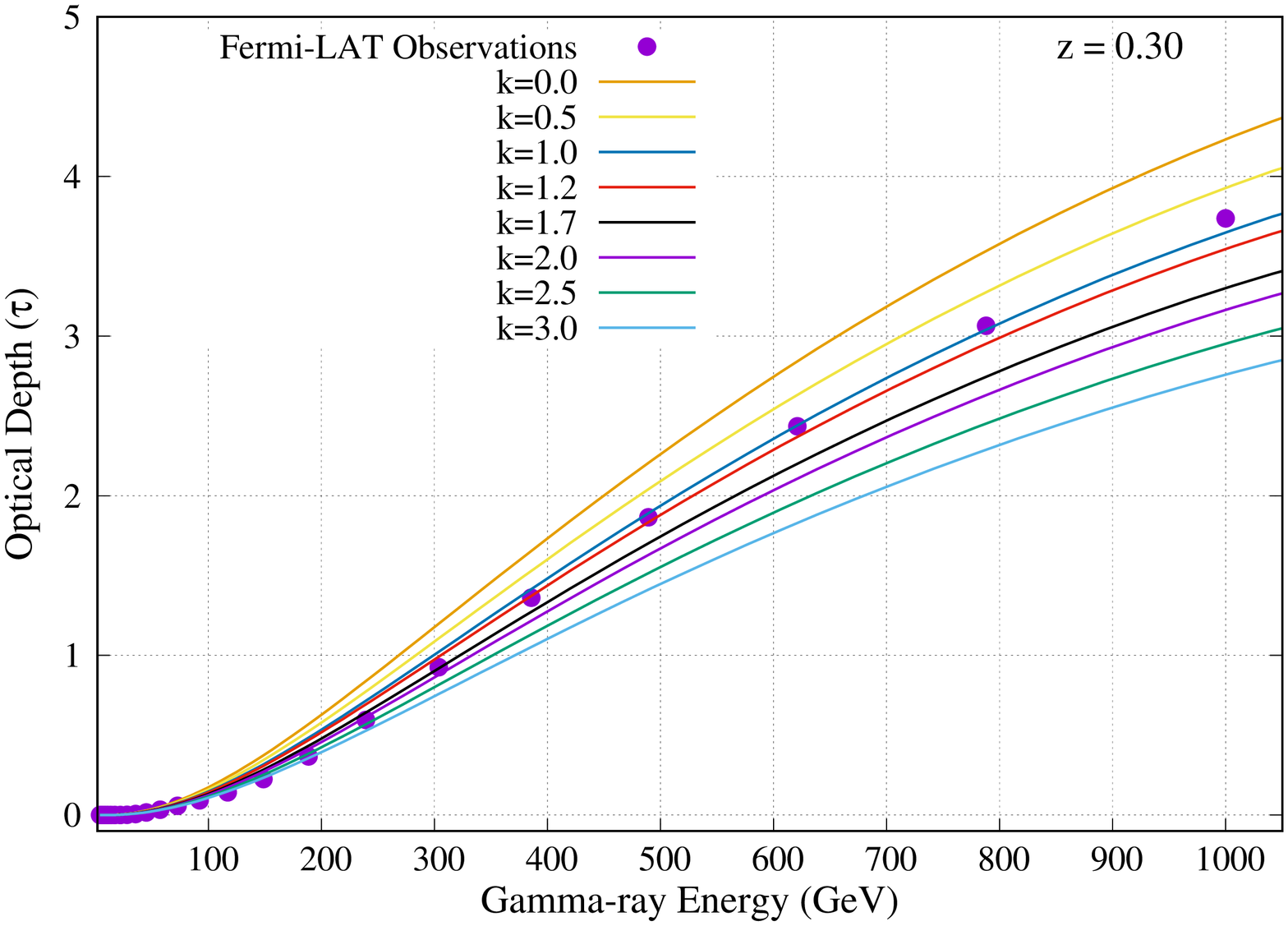}
\includegraphics*[width=0.32\columnwidth,angle=-90]{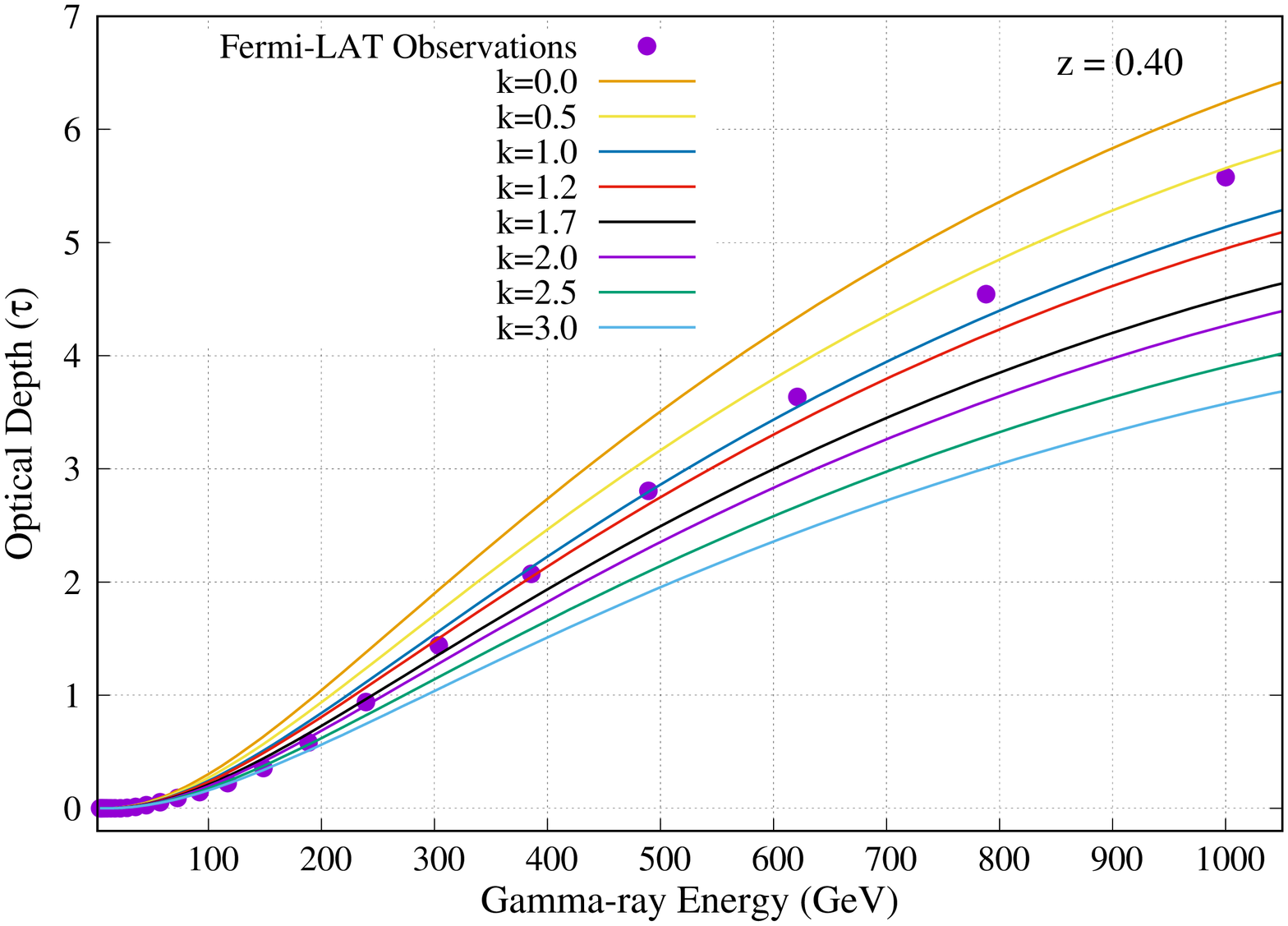}
\includegraphics*[width=0.32\columnwidth,angle=-90]{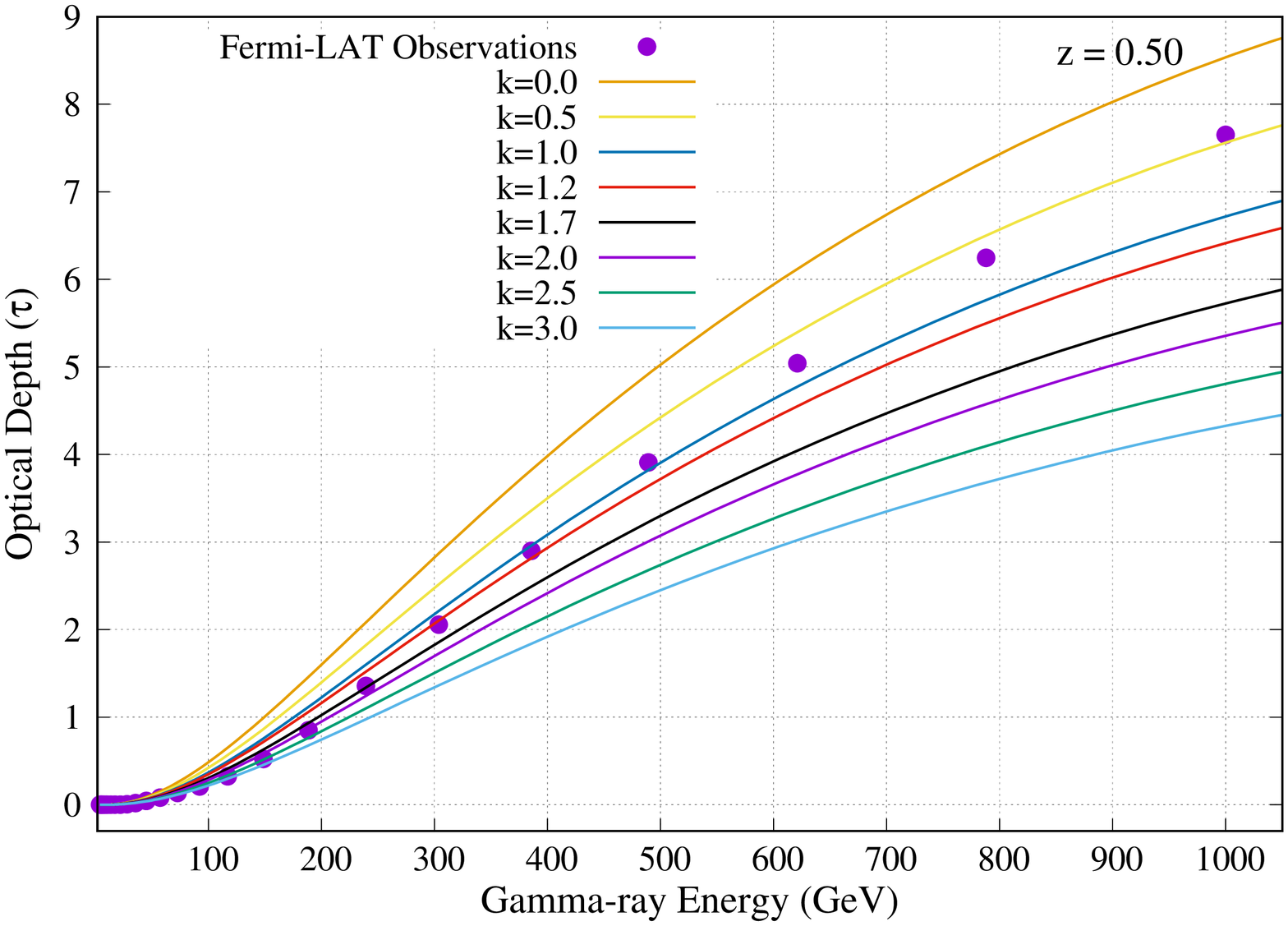}
\includegraphics*[width=0.32\columnwidth,angle=-90]{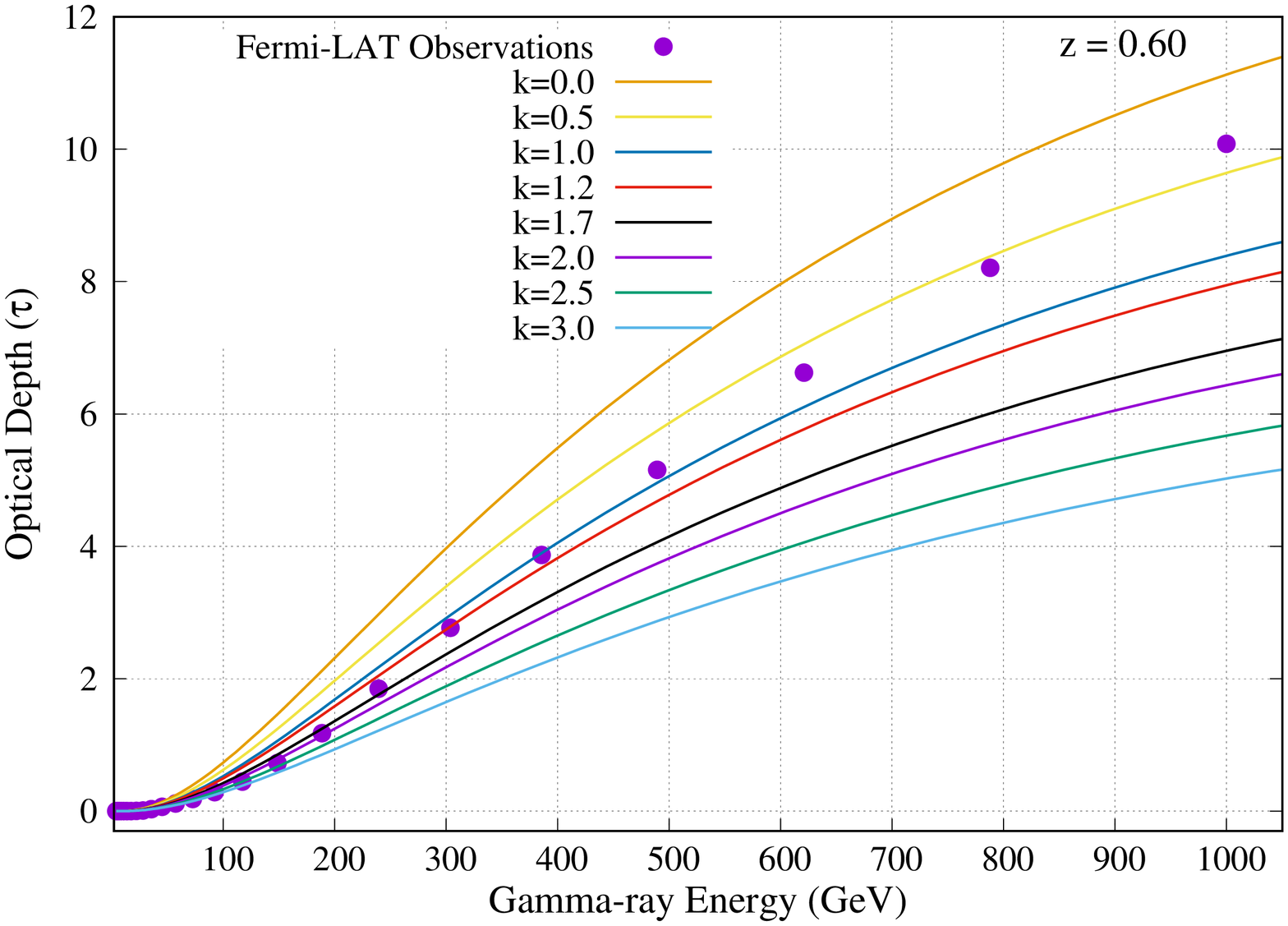}
\includegraphics*[width=0.32\columnwidth,angle=-90]{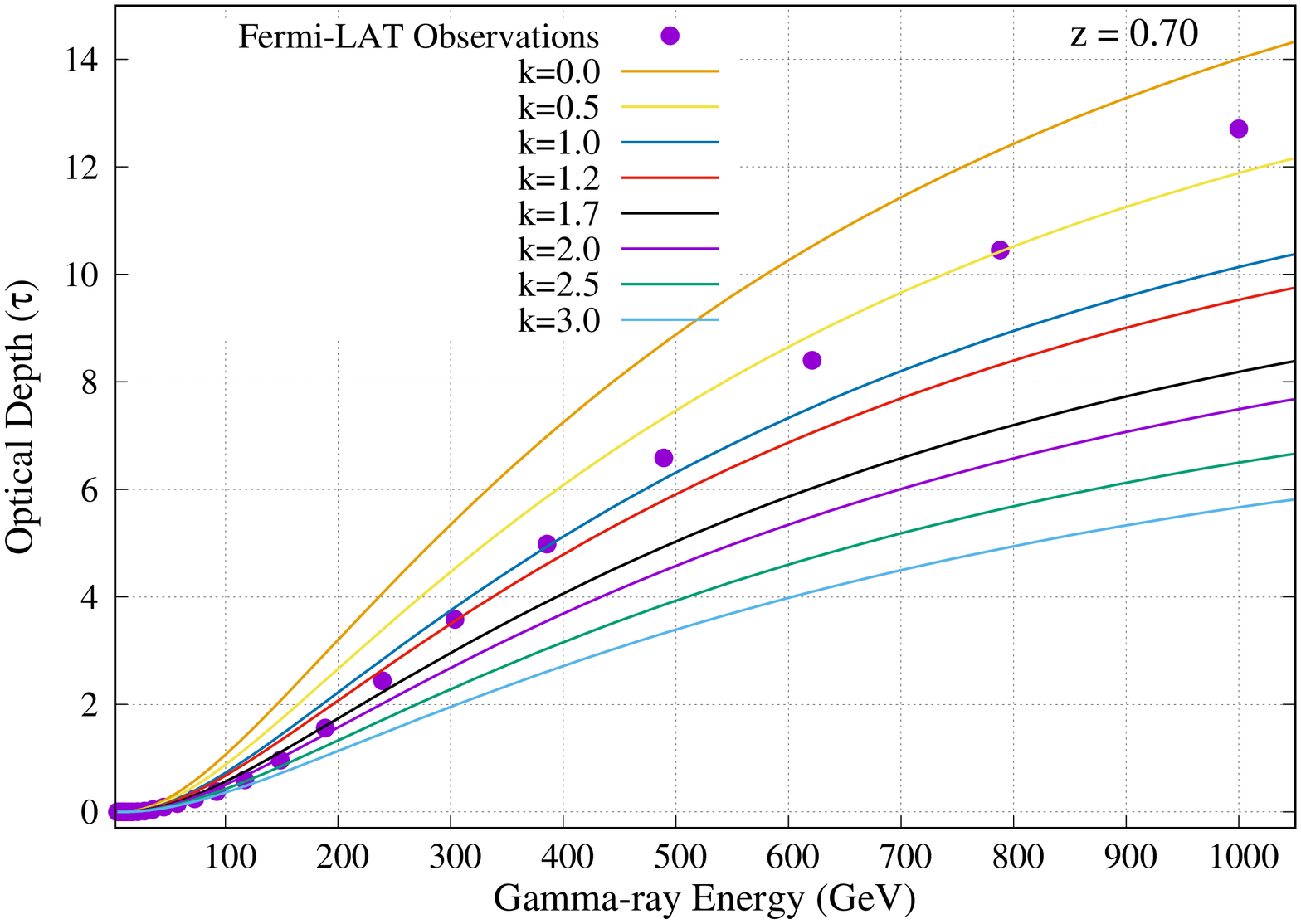}
\includegraphics*[width=0.32\columnwidth,angle=-90]{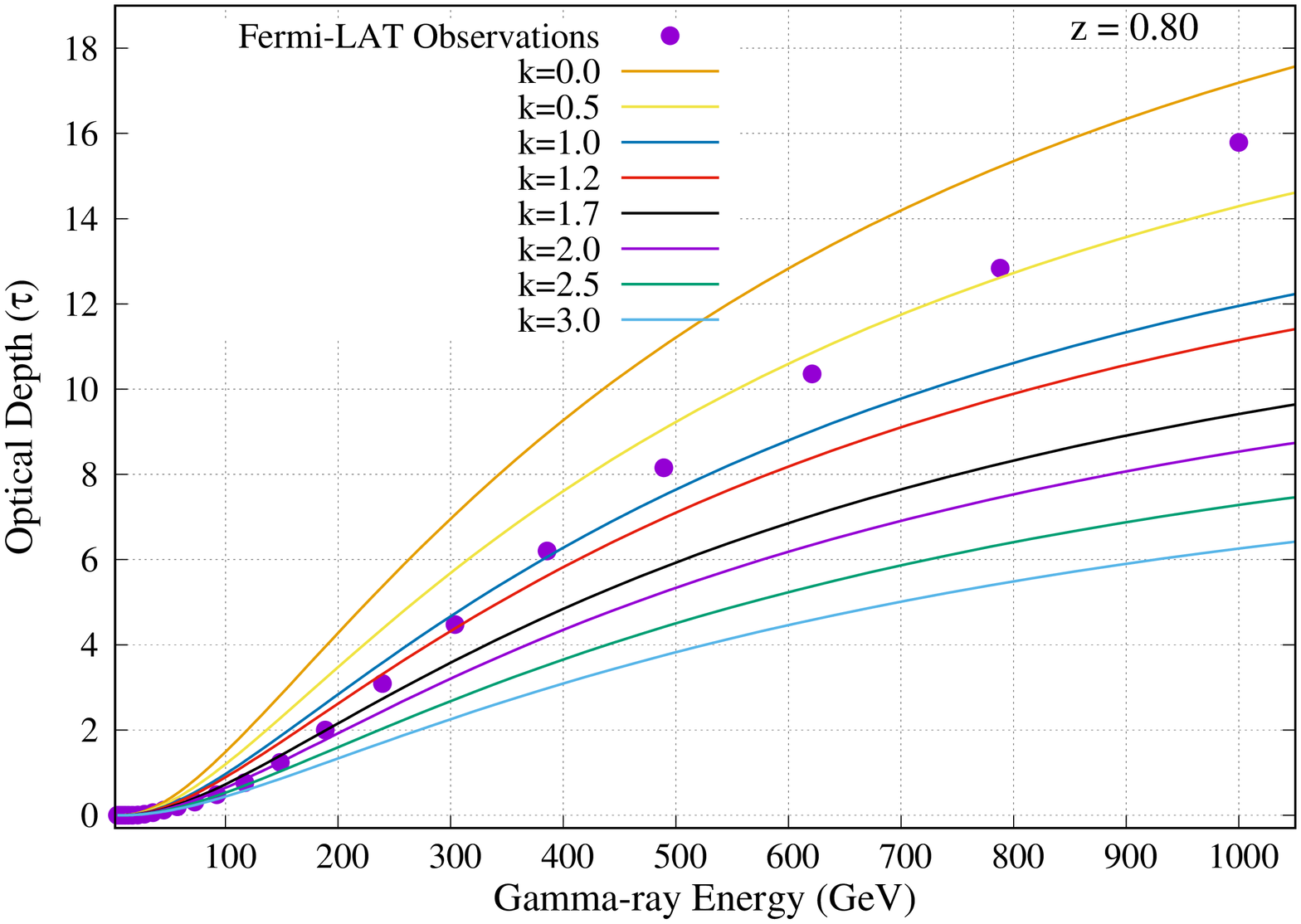}
\includegraphics*[width=0.32\columnwidth,angle=-90]{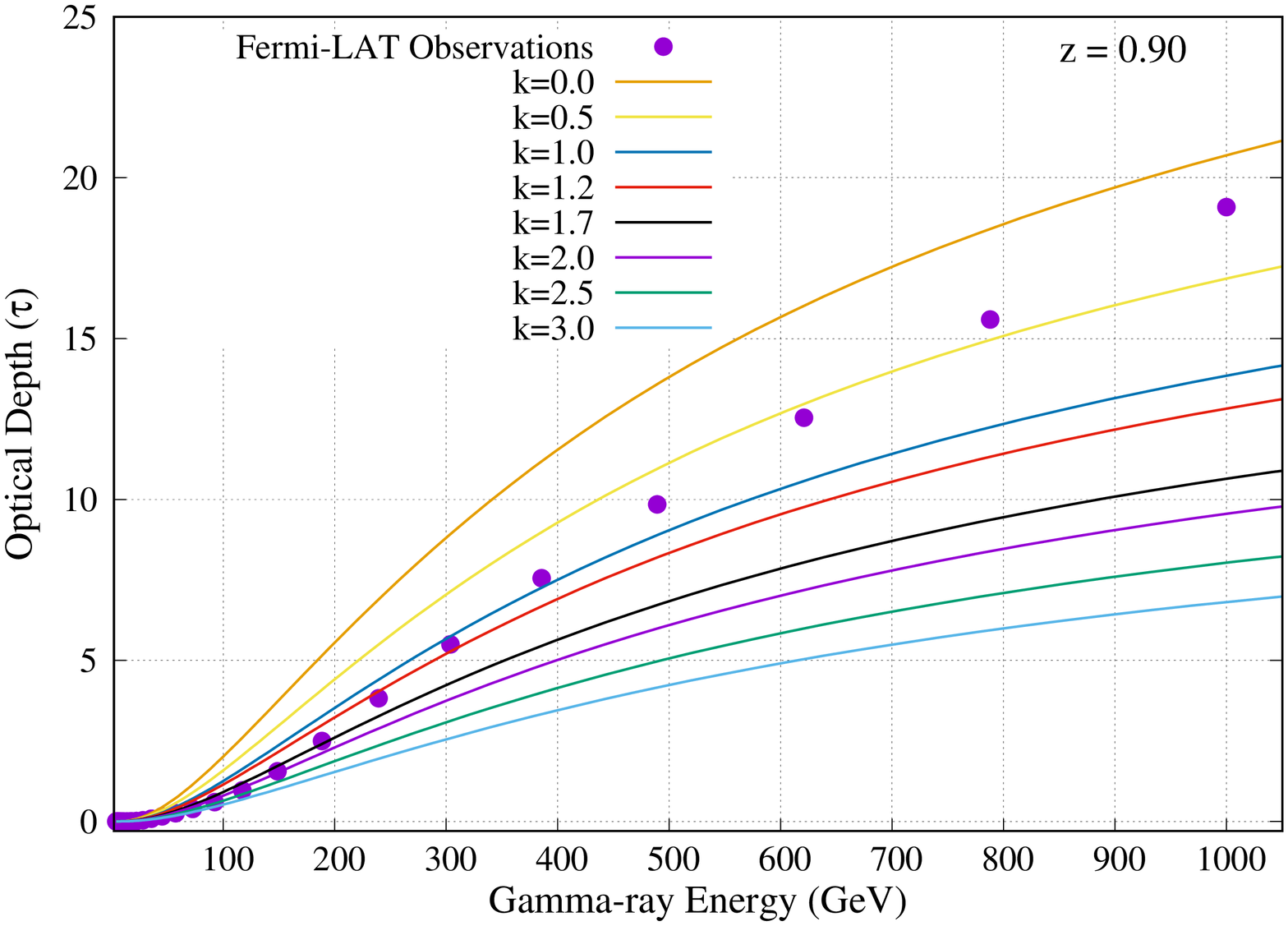}
\includegraphics*[width=0.32\columnwidth,angle=-90]{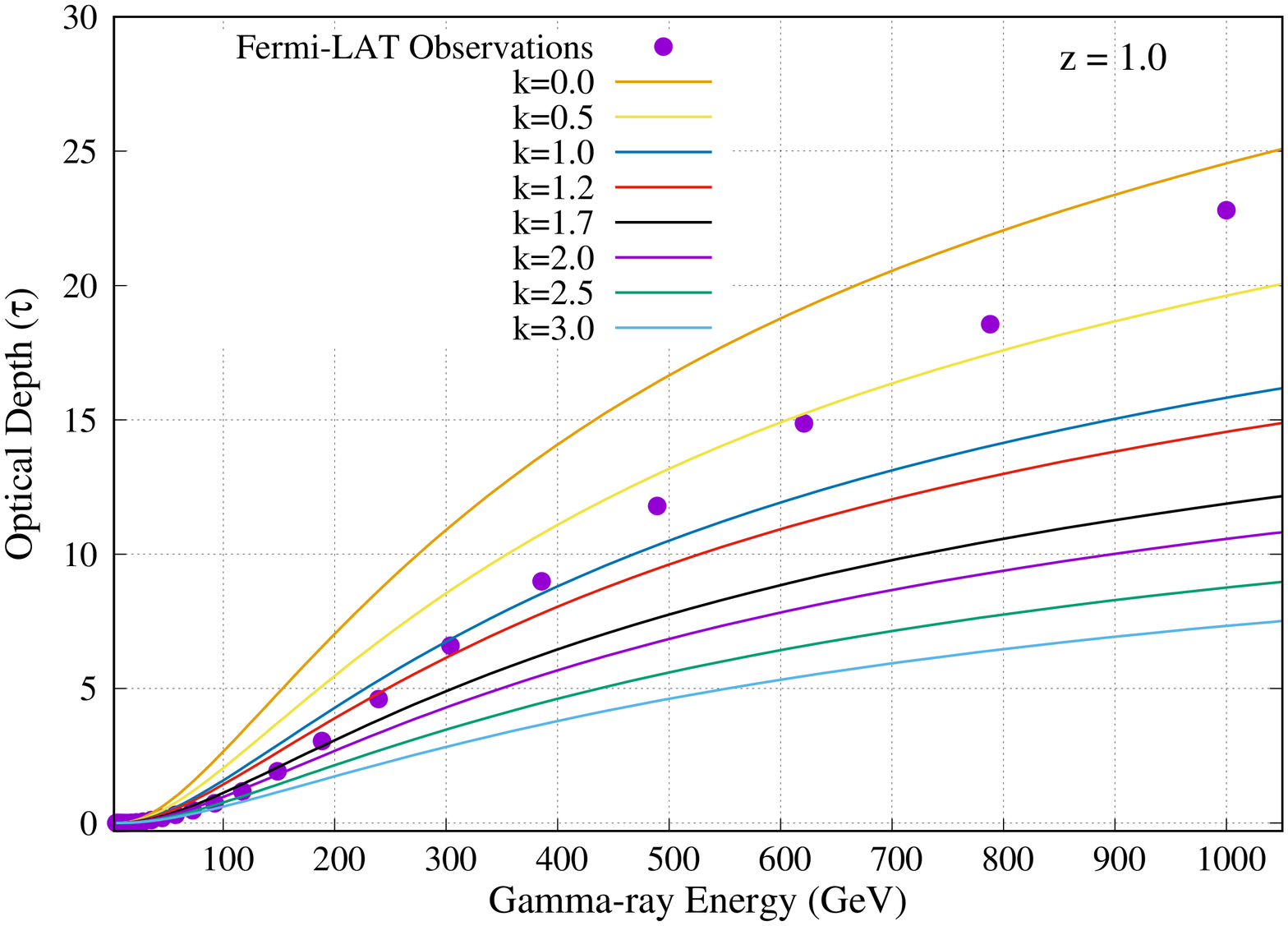}
\caption{Same as Figure \ref{od-Finke} for  the Dom\'inguez et al. (2011) EBL model}
\label{od-Doming}
\end{figure}

\begin{figure}[tb]
\includegraphics[width=0.7\columnwidth,angle=-90]{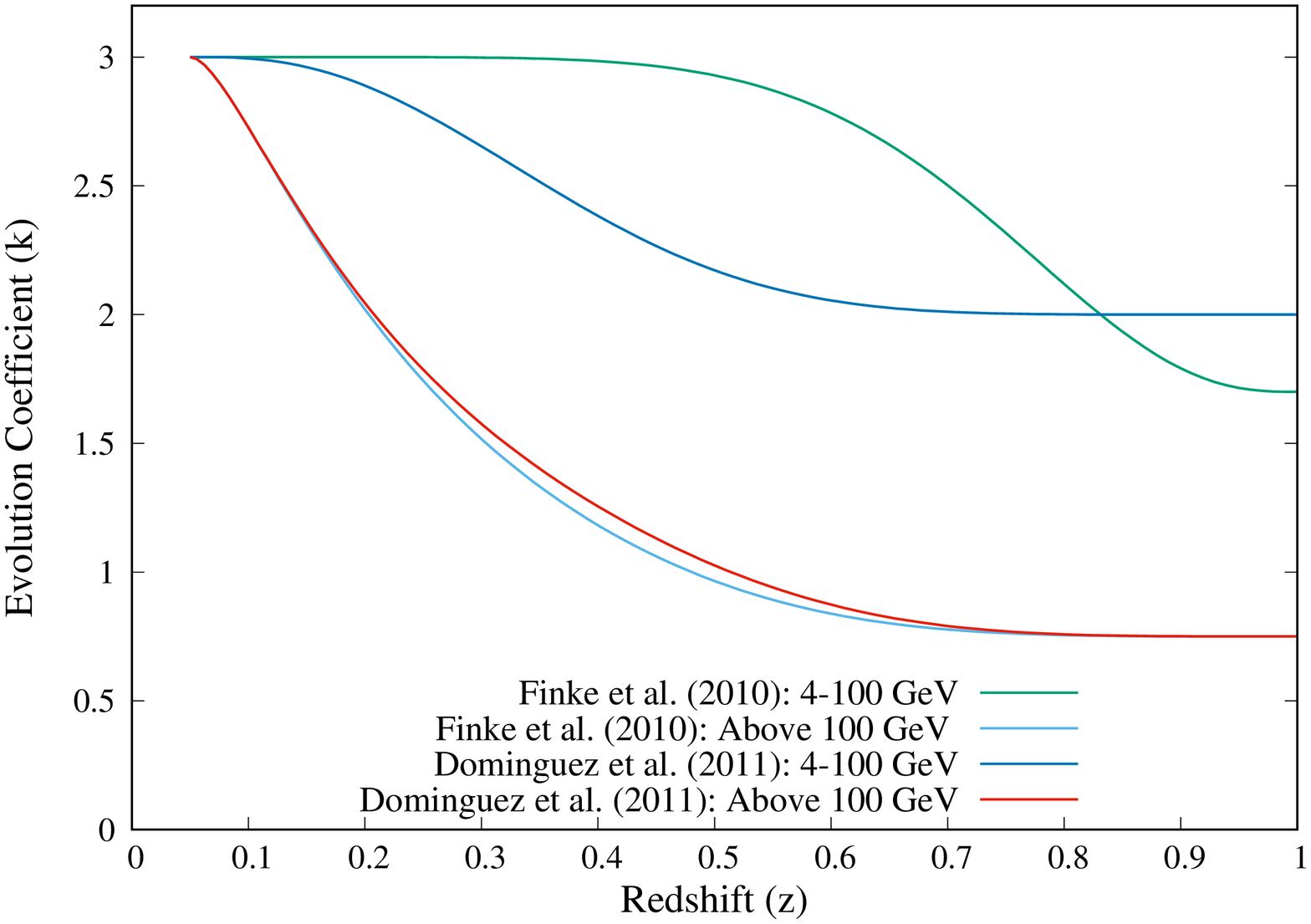}
\caption{Evolution coefficient $k$ as a function of $z$ for a good agreement between the optical depth values computed 
	 using the two EBL models and estimates from the \emph{Fermi}-LAT observations in two energy bands: 4-100 GeV and 
	 above 100 GeV (up to 1 TeV)}
\label{k-redshift}
\end{figure}

\section{$\gamma$-ray Propagation in \emph{Fermi}-Era}\label{s:gamma-prop}
The radiation field of the EBL behaves as a dominant source of the opacity for the high energy $\gamma$-ray photons travelling 
over the cosmological distances from the source towards the Earth. Photons in a $\gamma$-ray beam emitted from a distant source 
are attenuated by the EBL photon field via photon-photon pair production. The underlying interaction can be expressed through 
the Breit-Wheeler process as \citep{Gould1966,Gould1967}
\begin{equation}\label{pp}
	\gamma~~+~~\gamma_{EBL}~\rightarrow~ e^-~~+~~e^+
\end{equation}  
From the theory of quantum electrodynamics, the above interaction is kinematically allowed if the following condition is satisfied 
by the energies of two photons in the center of mass frame:
\begin{equation}\label{cm-ene}
	E_0~\epsilon_0~=~\frac{2 E_e^2}{(1 + z)^2 (1 - cos \rm \theta)}
\end{equation}
where $E_0$ is the observed energy of the $\gamma$-ray photon emitted from a source at redshift $z$, $\theta$ is the angle between 
the momenta of the $\gamma$-ray and the EBL photons, and $E_e$ is the total energy of electron (also the positron) produced in the 
pair creation (Equation \ref{pp}). The total scattering cross-section for the Breit-Wheeler process is given by \citep{Breit1934}
\begin{equation}\label{pp-cs}
	\sigma (\beta)~=~\frac{3\sigma_T}{16}(1-\beta^{2})\left[(3-\beta^{4})\rm ln\left(\frac{1+\beta}{1-\beta}\right) - 2\beta(2-\beta^{2})\right]
\end{equation}
where $\sigma_T$ is the Thomson cross section and $\beta$ is a parameter defined as 
\begin{equation}\label{beta}
	\beta (E_0, \epsilon_0, \theta, z)~=~\sqrt{1 - \frac{2m_{e}^{2}c^4}{E_0\epsilon_0 (1 + z)^2 (1-cos \rm \theta)}}
\end{equation} 
with $m_e c^2 (= 0.511 MeV)$ being the rest mass energy of the electron. The pair production cross section given by Equation \ref{pp-cs}  
has a peak value of $1.7 \times 10^{-25}~cm^2$ at $\beta = 0.70$ \citep{Gould1967}. This corresponds to the relation (from 
Equation \ref{beta}) 
\begin{equation}\label{ebl-peak}
            \epsilon_0~=~\frac{4 (m_e c^2)^2}{E_0 (1 + z)^2 (1 - cos \rm \theta)}
\end{equation}
This is the observed energy of the EBL photons which are most likely responsible for the pair production in the $\gamma-\gamma$ interaction. 
Attenuation of the $\gamma$-ray photons due to the interaction with the low energy background photons via pair creation is characterized by 
the optical depth ($\tau$) which strongly depends on the energy of the $\gamma$-ray photon ($E_0$), redshift of the $\gamma$-ray source 
($z_s$) and the proper number density of the EBL photons (n($\epsilon$,z)). The EBL optical depth to the $\gamma$-ray photons is computed as 
\begin{equation}\label{opacity}
	\tau(E_0,z_{s})=\int\limits_{0}^{z_{s}}\left(\frac{dl}{dz}\right)dz
	              \int\limits_{0}^{\pi}\left(\frac{1-\rm cos \rm \theta}{2}\right) \rm sin \rm \theta d \rm \theta
                      \int\limits_{\epsilon_{th}}^{\infty}n(\epsilon_0,0) (1 + z)^{3-k}
		      \sigma(\beta)d\epsilon
\end{equation}
where $\epsilon_{th}$ is the threshold energy of the EBL photon for the pair production, $\frac{dl}{dz}$ is the cosmological line element, 
and $\epsilon_0$ and $\epsilon$ are the EBL photon energies as defined under Equation \ref{ebl-den}. From Equation \ref{cm-ene}, we can 
write 
\begin{equation}\label{ebl-th}
            \epsilon_{th}~=~\frac{2 (m_e c^2)^2}{E_0 (1 + z)^2 (1 - cos \rm \theta)}
\end{equation} 
From Equations \ref{ebl-peak} and \ref{ebl-th}, it is evident that the EBL photons in the energy range $\approx$ 10$^{-3}$-10$^2$ eV 
play leading role in the absorption of the high energy $\gamma$-ray photons travelling over the cosmological distances with energies 
above 10 GeV. Attenuation due to the EBL strongly limits the propagation of the high energy $\gamma$-ray photons in the intergalactic 
space. The distance travelled by a $\gamma$-ray photon of energy $E_0$ corresponding to the redshift $z$ for which $\tau(E_0,z) = 1$, 
is referred to as the \emph{Gamma Ray Horizon} \citep{Fazio1970}. In the observational cosmology, the gamma ray horizon provides an 
estimate of the transparency of the Universe to the high energy photons. From radiative transfer theory, the gamma ray horizon predicts 
a redshift $z_s$ of a source for which the emitted $\gamma$-ray flux is attenuated by a factor $1/e$ for each observed energy $E_0$. 
Therefore, the sources beyond the gamma ray horizon will become progressively invisible. For head on encounter ($\theta = \pi$) between 
the $\gamma$-ray and the EBL photons, the interaction cross section for the pair production maximizes at redshift $z_{max}$ along the 
line of propagation, which is given by (rearranging Equation \ref{cm-ene} for $E_e = m_e c^2$)
\begin{equation}
	z_{max}~=~2.0\left(\frac{2~\rm eV}{\epsilon_0}\right)^{1/2}\left(\frac{30~\rm GeV}{E_0}\right)^{1/2}~-~1
\end{equation}
The EBL absorption feature has been observed in the $\gamma$-ray spectra of a sample of blazars in the energy range $E_0 =$ 1 - 500 GeV out 
to a redshift of $z~\sim~1.6$ detected by the \emph{Fermi}-LAT \citep{Ackermann2012}. This has also allowed to estimate the EBL intensity 
in the optical and UV wavebands by extracting the collective absorption effects on the $\gamma$-ray spectra of the blazars at different 
redshifts. Further, detection of the EBL attenuation in the spectra of a large sample of the active galaxies up to a redshift of  
$z \sim 3.1$ and one gamma-ray burst by the \emph{Fermi}-LAT observations in the energy range $E_0 =$ 10 - 1000 GeV, has allowed to 
determine the star formation history of the Universe up to $z \sim 6$ \citep{Abdollahi2018}. The \emph{Fermi}-LAT provides an excellent 
coverage of the whole $\gamma$-ray sky in wide energy range above 100 MeV. Recent measurements of the gamma ray horizon and highest energy of 
photons observed from a large sample of the blazars up to a redshift $z \le 1$ are depicted  in Figure \ref{lat-horizon} from 
the \emph{Fermi}-LAT observations \citep{Abdollahi2018}. We observe that the local Universe ($z \le 1$) is transparent to the high energy 
$\gamma$-ray photons with energies up to 500 GeV. Interestingly, the MAGIC telescopes have detected the significant $\gamma$-ray emissions 
in the energy bands 40-250 GeV and 65-175 GeV from the two most distant blazars at $z = 0.939$ \citep{Ahnen2015} and 
$z = 0.944$ \citep{Ahnen2016b} respectively. The VERITAS telescopes also reported the $\gamma$-ray emission up to 200 GeV from the blazar detected 
by the MAGIC telescopes at $z = 0.939$ \citep{Abeysekara2015}. These observations represent the most distant blazars detected to date and have 
significantly expanded the gamma ray horizon for the ground-based $\gamma$-ray telescopes. The highest energy of photons ($\sim$ 200 GeV) detected 
from these sources place stronger constraints on the gamma ray horizon from the \emph{Fermi}-LAT observations as shown in Figure \ref{lat-horizon}. 
It is obvious from Figure \ref{lat-horizon} that the Universe is transparent to the $\gamma$-ray photons with energy above 200 GeV emitted from a 
source at $z \sim 0.9$. Therefore, the $\gamma$-ray observations in the GeV energy band can be used as a powerful tool to probe the EBL in 
the local Universe.
\par
In the present work, we have assumed a flat $\Lambda$CDM cosmology with $\Omega_\Lambda = 0.7$, $\Omega_m = 0.3$ 
and  $H_0 = 70$ km s$^{-1}$ Mpc$^{-1}$. The cosmological line element for the propagation of the $\gamma$-ray photons in 
the flat $\Lambda$CDM cosmology is expressed as 
\begin{equation}
		\frac{dl}{dz}=\frac{c}{H_{0}}\frac{1}{(1+z)\sqrt{\Omega_{\Lambda}+\Omega_{m}(1+z)^{3}}}
\end{equation}

\section{Results and Discussion}\label{s:res-disc}
We aim to probe the cosmological evolution of the EBL photons in the local Universe ($z \le 1$) using the gamma ray horizon 
(Figure \ref{lat-horizon}) obtained from the \emph{Fermi}-LAT observations \citep{Abdollahi2018}. By definition, the gamma 
ray horizon represents a combination of $E_0$ and $z$ corresponding to $\tau(E_0,z) = 1$. We have selected such $E_0$ and $z$ 
combinations from Figure \ref{lat-horizon} (orange curve) and estimated $\tau(E_0,z) = 1$ using Equation \ref{opacity} by 
varying the evolution coefficient $k$ and assuming that the density of the EBL photons at the present epoch is described 
by the two models shown in Figure \ref{ebl-sed}. The variation of $k$ as a function of $z$ in the local Universe for the 
two EBL models is presented in Figure \ref{k-hor-opt}. We observe that $k$ strongly depends on the $z$ values for 
the observed $\gamma$-ray energies $E_0$ in the range 100- 500 GeV over the redshift range of $z =$0.2-1. For the Finke et al. (2010) model, 
$k$ increases from 3.0 to 3.5 corresponding to $z=0.2$ and 0.3 respectively and subsequently decreases to a minimum value 
of $\sim$ 2.0 at $z=1.0$ for the gamma ray horizon of photons in the energy range 100-500 GeV. Similarly, the value of $k$ first 
increases from 2.5 (at $z=0.2$) to 3.0 (at  $z=0.3$) followed by a rapid decrease to a value $\sim$ 2.0 at $z=1.0$ in case of 
the Dom\'inguez et al. (2011) model. This implies that the gamma ray horizon from the \emph{Fermi}-LAT observations for photons 
in the energy range $\approx$ 100-500 GeV suggests nearly similar evolution of the EBL photon density for the two EBL models 
employed in this study and predicts a value of $k$ between $\sim$ 2.0 and $\sim$ 3.0 in the local Universe $z \le 1$.

\par
We further estimate the optical depth values for the high energy $\gamma$-ray photons in the energy range 4 GeV to 1 TeV at a 
given source redshift ($z_s =$ 0.05,0.1,0.2,0.3,0.4,0.5,0.6,0.7,\\0.8,0.9,1.0) for different values of $k$ ranging between 
0.0 - 3.0 using Equation \ref{opacity} corresponding to the two EBL models. A comparison of the optical depth values derived 
using the EBL model reported by \cite{Abdollahi2018} from the \emph{Fermi}-LAT observations\footnote{https://figshare.com/s/14f943002230d69a4afd} 
with the corresponding estimates for the Finke et al. (2010) and Dom\'inguez et al. (2011) EBL models is shown in Figure \ref{od-Finke} and 
Figure \ref{od-Doming} respectively. It is obvious from both the figures (\ref{od-Finke} \& \ref{od-Doming}) that a close matching between 
the two opacity values is observed for  different values of $k$ at various redshifts in the different energy range of the $\gamma$-ray photons. 
The variation of $k$ with $z$ for a close matching between the computed and measured opacity of the Universe to the high energy $\gamma$-ray 
photons in the two energy bands 4-100 GeV and above 100 GeV (up to 1 TeV) is reported in Figure \ref{k-redshift}. For $E_0 \le$ 100 GeV, the 
optical depth values are consistent with each other for $k = 3$ up to $z < 0.2$ (low redshift) for the two EBL models. Beyond redshift $z > 0.2$, 
the value of $k$ decreases from 3.0 to 1.7 and 2.0 corresponding to the Finke et al. (2010) and Dom\'inguez et al. (2011) models respectively 
at $z = 1.0$ in the local Universe (Figure \ref{k-redshift}). This suggests that the evolution coefficient shows completely different behaviour 
in the local Universe for the two EBL models and the values of $k$ can be inferred in the range 3.0-1.7 and 3.0-2.0 for the Finke et al. (2010) 
and Dom\'inguez et al. (2011) models respectively. Above 100 GeV, the agreement between the derived optical depth values for the two EBL models 
and the \emph{Fermi}-LAT estimates is obtained for $k = 3.0$ up to $z < 0.1$. Beyond this redshift, the value of $k$ is observed to decrease 
very rapidly with increasing $z$ and attains a common value of $\sim$ 0.75 at  $z = 1.0$ (Figure \ref{k-redshift}) for both the EBL models. This 
indicates that the variations in the value of $k$ derived from the gamma ray horizon are broadly consistent with the inferences from the 
comparison of the optical depth estimates. For both the EBL models, $k = 3$ suggests no cosmological evolution of the EBL photon density at 
redshifts below 0.1. The value of $k$ decreases with increasing $z$ at higher redshifts beyond $z \ge 0.1$ for the $\gamma$-ray energies 
up to 1 TeV. 
\par
The gamma ray horizon of the Universe to the TeV $\gamma$ rays is limited to $z < 1$. Recent observations of the most distant blazars 
at $z = 0.9$ with the MAGIC and VERITAS telescopes are limited to highest energies up to $\sim$ 200 GeV \citep{Abeysekara2015,Ahnen2015,Ahnen2016b}. 
The observed $\gamma$-ray spectra of these sources are very steep with power law spectral indices $>$ 3.5. However, after corrections for the 
expected EBL absorption, their intrinsic spectra are found to be very hard with the power law spectral indices $<$ 1.5. From the standard 
scenario for high energy $\gamma$-ray emission from blazars, the intrinsic spectra cannot be harder than 1.5 \citep{Aharonian2006}. However, 
the current statistics of the $\gamma$-ray observations of the blazars with the ground-based telescopes do not allow any robust conclusion 
regarding the intrinsic $\gamma$-ray spectra above 1 TeV for sources at $z \sim 1$. The gamma ray horizon predicted by a model-independent 
EBL measurement with the H.E.S.S. array is compatible with the predictions from the Finke et al. (2010) and Dom\'inguez et al. (2011) models, 
but the sensitivity of this approach is limited due to the consideration of systematic uncertainties in the horizon envelope up to $z \sim 0.3$ and 
energy less than 1 TeV \citep{Abdalla2017}.
\par
From the literature, the evolution proposed by Raue \& Mazin (2008), $k = 1.2$, leads to a significant agreement for 
redshift up to $z \sim$ 0.7 provided the EBL photon density at $z = 0$ is described by a generic model which is in 
compliance with the lower and upper EBL limits. The present epoch EBL density predicted by this generic model is just above 
the lower limits derived from the galaxy source counts and the SED of the EBL simply represents a fit to the existing limits and not 
a complete theoretical model \citep{Raue2008}. A template evolution with $k = 1.7$ for another EBL model in \citep{Gilmore2012} is found to 
be in good agreement with the $\gamma$-ray observations up to redshift $z =$ 0.6 \citep{Biteau2015}. However, optical depths are underestimated 
by the template evolution with $k = 2.2$ at higher redshifts. This model is based on the semi-analytical approach for simulating the galaxy 
formation and evolution involving complex physical processes in the EBL emission \citep{Gilmore2012}. These values are broadly consistent 
with the $k$ values obtained in the present work for the two EBL models. The EBL models employed in this study do not require any complex stellar 
structure code or semi-analytical models of the galaxy formation. The star formation history determined by the \emph{Fermi}-LAT observations out 
to a redshift of $z \sim$ 5 is in agreement with the independent measurements of the galaxy counts with a peak at $z \sim$ 2 \citep{Abdollahi2018}.

\section{Conclusions}\label{s:conclu}
The cosmological evolution of the EBL photon number density suggests that the EBL does not represent instantaneously produced 
background photons. The UV/optical and IR photons contributing to the broadband SED of the EBL are built up slowly over the 
history of the Universe from the epoch of recombination to the present epoch. Therefore, the number density of the EBL photons 
at the present epoch ($z = 0$) is scaled by a factor $(1 + z)^{3 - k}$, where value of the evolution coefficient $k$ can be 
tuned as summarized below: 

\begin{itemize}
\item Cosmological evolution of the EBL photon density in the local Universe cannot be described by a unique value of the 
	evolution coefficient $k$. The value of $k$ varies between $k = 3$ and $k = 0.75$ corresponding to the low ($z \le$ 0.1) and 
	high redshifts ($z \sim$ 1) respectively. 

\item $k = 3$ suggests no evolution of the EBL photon density at low redshifts and is compatible with the transparency of the Universe 
	to the $\gamma$-rays with energy below 100 GeV.  

\item $k = 0$ represents a simple cosmological dilution of the EBL photon field due to expansion of the Universe. However, the present 
	study suggests $k~\ge~0.75$ in the local Universe ($z \sim$ 1) for the high energy $\gamma$-ray photons with energy above 100 GeV 
	(up to 1 TeV).

\item For the EBL photon density described by Finke et al. (2010) and Dom\'inguez et al. (2011) at the present epoch, the cosmological evolution 
	can be broadly described by a mean value of $k$ in the range 0.75 - 3 for the observed $\gamma$-ray energies in the range 4 GeV-1 TeV.

\item The value of $k = 1.7$ widely used in the literature is consistent with the results derived in the present study at $z \sim$ 0.6-0.7  for 
	the gamma ray horizon predicted  by the Dom\'inguez et al. (2011) model in the GeV energy regime.  
              
\end{itemize}
The complex behaviour of evolution coefficient $k$ as a function of redshift can be further addressed significantly by the 
new-generation ground-based Cherenkov Telescope Array (CTA) observatory \citep{Acharyya2019}. The CTA observations over a 
wide energy range are expected to explore the effect of the EBL on the $\gamma$-ray propagation up to a redshift beyond $z = $ 1.

\acknowledgments
Authors thank the anonymous reviewers for their important suggestions and critical 
comments that greatly helped to improve the manuscript.



\begin{thebibliography}{}
	\bibitem[Abdalla et al. (2017)]{Abdalla2017} Abdalla, H., et al.: Astron. Astrophys. {\bf 606}, 59 (2017)	
	\bibitem[Abdollahi et al. (2018)]{Abdollahi2018} Abdollahi, S., et al.: Science {\bf 362}, 1031 (2018)    
	\bibitem[Abeysekara et al. (2015)]{Abeysekara2015} Abeysekara, A.U., et al.: Astrophys. J. Lett. {\bf 815}, 22 (2015)
	\bibitem[Abeysekara et al. (2019)]{Abeysekara2019} Abeysekara, A.U., et al.: Astrophys. J. {\bf 885}, 150 (2019)
	\bibitem[Abramowski et al. (2013)]{Abramowski2013} Abramowski, A., et al.: Astron. Astrophys. {\bf 550}, 4 (2013)
	\bibitem[Acciari et al. (2019)]{Acciari2019} Acciari, V.A., et al.: Mon. Not. R. Astron. Soc. {\bf 486}, 4233 (2019)
	\bibitem[Acharyya et al. (2019)]{Acharyya2019} Acharyya, A., et al.: Astroparticle Physics {\bf 111}, 35 (2019)
	\bibitem[Ackermann et al. (2012)]{Ackermann2012} Ackermann, M., et al.: Science {\bf 338}, 1190 (2012)
	\bibitem[Aharonian et al. (2006)]{Aharonian2006} Aharonian, F., et al.: Nature {\bf 440}, 1018 (2006)
	\bibitem[Aharonian et al. (2007)]{Aharonian2007} Aharonian, F., et al.: Astron. Astrophys. {\bf 475}, 9 (2007)
	\bibitem[Ahnen et al. (2015)]{Ahnen2015} Ahnen, M.L., et al.: Astrophys. J. Lett. {\bf 815}, 23 (2015)
	\bibitem[Ahnen et al. (2016a)]{Ahnen2016a} Ahnen, M.L., et al.: Astron. Astrophys. {\bf 595}, 98 (2016a)
	\bibitem[Ahnen et al. (2016b)]{Ahnen2016b} Ahnen, M.L., et al.: Astron. Astrophys. {\bf 590}, 24 (2016b)
	\bibitem[Biteau \& Williams (2015)]{Biteau2015} Biteau, J., Williams, D.A.: Astrophys. J. {\bf 812}, 60 (2015)
	\bibitem[Breit \&  Wheeler (1934)]{Breit1934} Breit, G., Wheeler, J.A.: Physical Review {\bf 46}, 1087 (1934)
	\bibitem[Cowley et al. (2019)]{Cowley2019} Cowley, W.I., et al.: Mon. Not. R. Astron. Soc. {\bf 487}, 3082 (2019)
	\bibitem[Desai et al. (2017)]{Desai2017} Desai, A., et al.: Astrophys. J. {\bf 850}, 73 (2017)
	\bibitem[Desai et al. (2019)]{Desai2019} Desai, A., et al.: Astrophys. J. Lett. {\bf 874}, 7 (2019)
	\bibitem[Dole et al. (2006)]{Dole2006} Dole, H., et al.: Astron. Astrophys. {\bf 451}, 417 (2006)
	\bibitem[Dom\'inguez et al. (2011)]{Dom2011} Dom\'inguez , A., et al.: Mon. Not. R. Astron. Soc. {\bf 410}, 2556 (2011)
	\bibitem[Driver et al. (2016)]{Driver2016} Driver, S.P., et al.: Astrophys. J. {\bf 827}, 108 (2016)
	\bibitem[Dwek \& Krennrich (2013)]{Dwek2013} Dwek, E., Krennrich, F.: Astroparticle Physics {\bf 43}, 112 (2013)
	\bibitem[Fan et al. (2006)]{Fan2006} Fan, X., et al.: Astron. J. {\bf 132}, 117 (2006)
	\bibitem[Fazio \& Stecker (1970)]{Fazio1970} Fazio, G. G., Stecker, F.W.: Nature {\bf 226}, 135 (1970)
	\bibitem[Finke et al. (2010)]{Finke2010} Finke, J.D., et al.: Astrophys. J. {\bf 712}, 238 (2010)
	\bibitem[Franceschini \& Rodighiero (2017)]{Franceschini2017} Franceschini, A., Rodighiero, G.: Astron. Astrophys. {\bf 603}, 34 (2017)
	\bibitem[Gilmore et al. (2009)]{Gilmore2009} Gilmore, R.C., et al.: Mon. Not. R. Astron. Soc. {\bf 399}, 1694 (2009)
	\bibitem[Gilmore et al. (2012)]{Gilmore2012} Gilmore, R.C., et al.: Mon. Not. R. Astron. Soc. {\bf 422}, 3189 (2012)
	\bibitem[Gould \& Schr\'eder (1966)]{Gould1966} Gould, R.J., Schr\'eder, G.P.: Phys. Rev. Lett. {\bf 16}, 252 (1966)
	\bibitem[Gould \& Schr\'eder (1967)]{Gould1967} Gould, R.J., Schr\'eder, G.P.: Physical Review {\bf 155}, 1404 (1967)
	\bibitem[Hauser et al. (1998)]{Hauser1998} Hauser, M.G., et al.: Astrophys. J. {\bf 508}, 25 (1998)
	\bibitem[Hauser \& Dwek (2001)]{Hauser2001} Hauser, M.G., Dwek, E.: Annu. Rev. Astron. Astrophys. {\bf 39}, 249 (2001)
	\bibitem[Keenan et al. (2010)]{Keenan2010} Keenan, R.C., et al.: Astrophys. J. {\bf 723}, 40 (2010)
	\bibitem[Khaire \& Srianand (2019)]{Khaire2019} Khaire, V., Srianand, R.: Mon. Not. R. Astron. Soc. {\bf 484}, 4174 (2019)
	\bibitem[Kneiske \& Dole (2010)]{Kneiske2010} Kneiske, T.M., Dole, H.: Astron. Astrophys. {\bf 515}, 19 (2010)
	\bibitem[Kneiske et al. (2002)]{Kneiske2002} Kneiske, T.M., et al.: Astron. Astrophys. {\bf 386}, 1 (2002)
	\bibitem[Madau \& Phinney (1996)]{Madau1996} Madau, P., Phinney, E.S.: Astrophys. J. {\bf 456}, 124 (1996)
	\bibitem[Madau \& Pozzetti (2000)]{Madau2000} Madau, P., Pozzetti, L.: Mon. Not. R. Astron. Soc. {\bf 312}, 9 (2000)
	\bibitem[Matsuura et al. (2017)]{Matsuura2017} Matsuura, S., et al.: Astrophys. J. {\bf 839}, 7 (2017)
	\bibitem[Mattila \& V\"ais\"anen (2019)]{Mattila2019} Mattila, K., V\"ais\"anen, P.: Contemporary Physics {\bf 60}, 23 (2019)
	\bibitem[Mazin \& Raue (2007)]{Mazin2007} Mazin, D., Raue, M.: Astron. Astrophys. {\bf 471}, 439 (2007)
	\bibitem[Meyer et al. (2012)]{Meyer2012} Meyer, M., et al.: Astron. Astrophys. {\bf 542}, 59 (2012)
	\bibitem[Perlmutter et al. (1999)]{Perlmutter1999} Perlmutter, S., et al.: Astrophys. J. {\bf 517}, 565 (1999)
	\bibitem[Raue \& Mazin (2008)]{Raue2008} Raue, M., Mazin, D.: International Journal of Modern Physics D {\bf 17}, 1515 (2008)
	\bibitem[Raue \& Meyer (2012)]{Raue2012} Raue, M., Meyer, M.: Mon. Not. R. Astron. Soc. {\bf 426}, 1097 (2012)
	\bibitem[Riess et al. (1998)]{Riess1998} Riess, A.G., et al.: Astron. J. {\bf 116}, 1009 (1998)
	\bibitem[Saldana-Lopez et al. (2020)]{Saldana2020} Saldana-Lopez, A., et al.: arXiv:2012.03035 (2020)
	\bibitem[Singh et al. (2014)]{Singh2014} Singh, K.K., et al.: New Astron. {\bf 27}, 34 (2014)
	\bibitem[Singh et al. (2019)]{Singh2019} Singh, K.K., et al.: Experimental Astronomy {\bf 48}, 297 (2019)
	\bibitem[Singh \& Meintjes (2020)]{Singh2020} Singh, K.K., Meintjes, P.J.: NRIAG Journal of Astronomy and Geophysics {\bf 9}, 309 (2020)
	\bibitem[Stecker et al. (1992)]{Stecker1992} Stecker, F.W., et al.: Astrophys. J. Lett. {\bf 390}, 49 (1992)
	\bibitem[Stecker et al. (2016)]{Stecker2016} Stecker, F.W., et al.: Astrophys. J. {\bf 827}, 6 (2016)
	\bibitem[Zemcov et al. (2017)]{Zemcov2017} Zemcov, M., et al.: Nature Communications {\bf 8}, 15003 (2017)
\end{thebibliography}
\end{document}